\documentclass[twocolumn,preprintnumbers,amsmath,amssymb,longbibliography]{revtex4-1}

\usepackage{graphicx}
\usepackage{dcolumn}
\usepackage{float} 
\usepackage{bbm}

\usepackage{setspace}

\begin{document}

\title{Discontinuities in driven spin-boson systems due to coherent destruction of tunneling: breakdown of the Floquet-Gibbs distribution}

\author{ Georg Engelhardt$^1$    }

\author{ Gloria Platero $^2$    }

\author{ Jianshu Cao$^{1,3} $}
\email{jianshu@mit.edu}

\affiliation{%
$^1$Beijing Computational Science Research Center, Beijing 100193, Peopleʼs Republic of China\\
$^2$Instituto de Ciencia de Materiales de Madrid, CSIC, 28049 Madrid, Spain\\
$^3$Department of Chemistry, Massachusetts Institute of Technology, 77 Massachusetts Avenue,
Cambridge, Massachusetts 02139, USA
}

\date{\today}

\pacs{
}

\begin{abstract}
We show that the probability distribution of the stationary state of a dissipative  ac-driven two-level system  exhibits  discontinuities, i.e. jumps,  for parameters at which coherent destruction of tunneling takes place. These discontinuities can be observed as jumps in the emission of the  Mollow triplet. The jumps  are the consequence of discontinuities in the transition rates, which we calculate numerically and analytically based on the secular Floquet-Redfield formalism.
\end{abstract}
\maketitle

\allowdisplaybreaks

\textbf{Introduction.}
Due to the high experimental control, periodic driving  has become a flexible tool for  quantum state manipulation with various applications, e.g., for topological matter, quantum phase transitions, quantum transport, and even-harmonic generation~\cite{Aidelsburger2013,Jotzu2014,Benito2014,Engelhardt2017a,
Bastidas2012,Engelhardt2016,Engelhardt2013,Liu2017a,
Platero2004,Bavli1993,Dakhnovskii1993,Graves1989,Mann2017,Komnik2016}.    
As a quantum system is never completely decoupled from its environment,  thermalization  finally leads to a relaxation towards a stationary state. Yet, very little is know about possible  stationary states of periodically-driven systems. In this letter, we report on exotic stationary states exhibiting probability jumps, which are related to two prominent effects, namely coherent destruction of tunneling (CDT)~\cite{Grossmann1991} and the Mollow triplet~\cite{Mollow1969}.

 In a recent article, Shirai et al. have discussed under which special conditions effective Floquet-Gibbs states arise~\cite{Shirai2015,Shirai2016}. The probabilities of  these periodic Floquet states, characteristic states of periodically-driven systems, are determined by their corresponding \textit{quasienergies} $\epsilon_{\lambda}$ in a Gibbs-like fashion, thus $p_{\lambda}\propto e^{-\beta \epsilon_{\lambda} }$. However, given the richness of quantum effects in periodically-driven systems and their experimental control, the resulting stationary states are not necessary Gibbs-Floquet states and can possibly exhibit intriguing features.

\begin{figure}[t]
\includegraphics[width=1\linewidth]{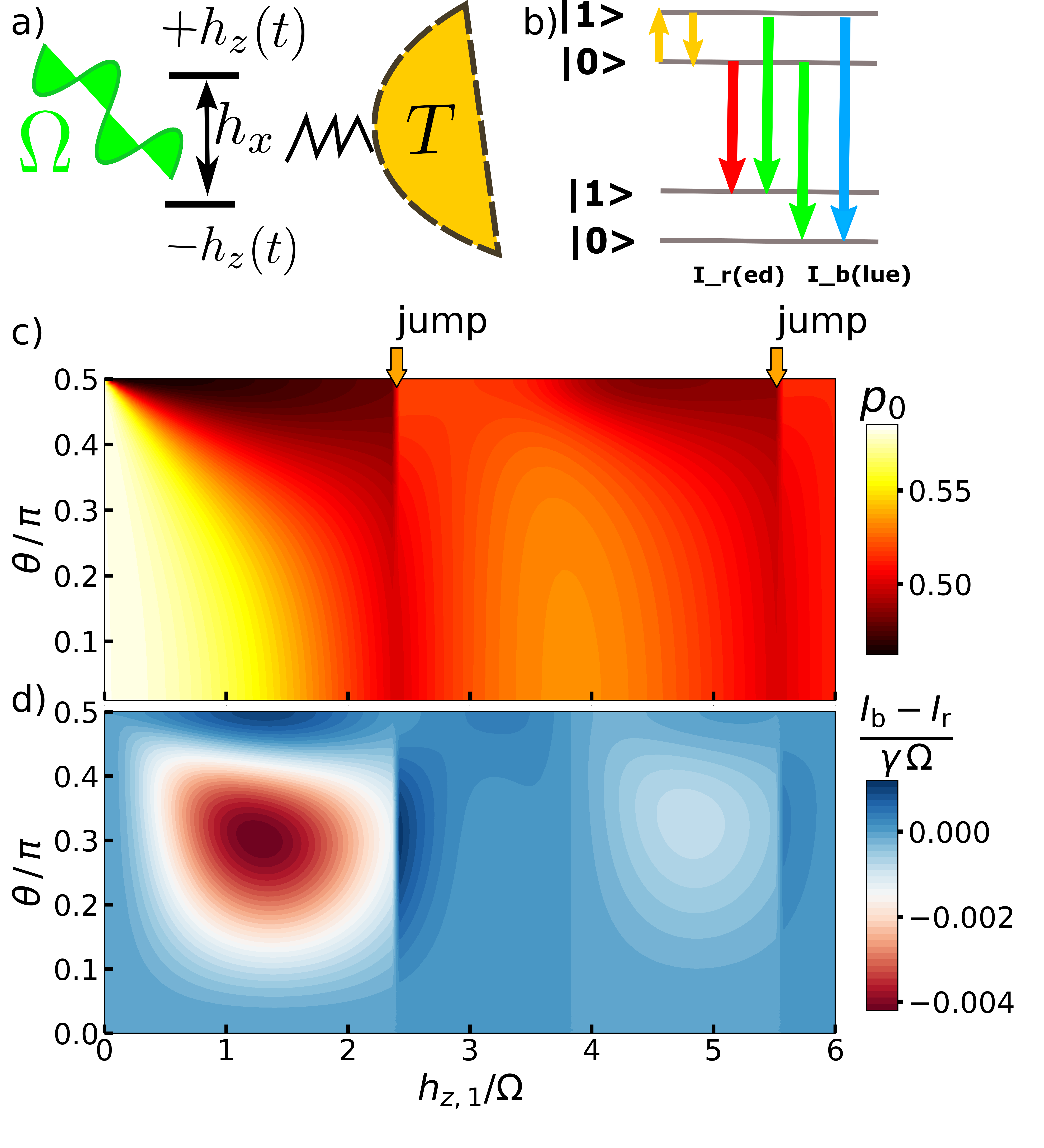}
\caption{ (a) The spin-boson model is driven with frequency $\Omega$. Due two the internal coherent dynamics, it emits phonons  with $\Omega,\Omega\pm \Delta$ (unshifted, blue shifted, red shifted). The corresponding transitions are depicted in (b). (c) shows the  probability of the  Floquet state for $\lambda=0$ as a function of the coupling angle $\theta$ and the  driving amplitude $h_{\rm z,1}$. (d) depicts the difference of the emitted  blue and red shifted phonons into the bath. The offset energy splitting $h_{-rm z,0}=0$, cut-off frequency $\omega_c = 10 h_{\rm x}$, driving frequency $\Omega=40 h_{\rm x}$ and temperature $ k_{\rm B}T= 3 h_{\rm x}$
are expressed in units of the tunneling amplitude $h_{\rm x}$.}
\label{fig:overview}
\end{figure}%

In  isolated ac-driven systems,  CDT is defined by an exact degeneracy of two quasienergies. Consequently, the dynamics in these levels is frozen as the effective Hamiltonian is zero~\cite{Frasca2003,Stockburger1999,Barata2000}. 
The emission  of an ac-driven two-level system ($\left|0\right>,\left|1\right>$)  coupled to an environment (sketched in Fig.~\ref{fig:overview}(a)) can exhibit a Mollow triplet as sketched in Fig.~\ref{fig:overview}(b)~\cite{Mollow1969,Pigeau2015,Yan2016}. Thereby, the  driving field with frequency $\Omega $ induces transitions between the states which are accompanied by emitting a photon (or phonon, which we focus here) with frequencies  $\Omega - \Delta$ ($\left|0\right>\rightarrow\left|1\right>$), $\Omega$ ($\left|0\right>\rightarrow\left|0\right>$ and $\left|1\right>\rightarrow\left|1\right>$), and $\Omega + \Delta$ ($\left|1\right>\rightarrow\left|0\right>$). Here, $\Delta$ denotes the  difference between two quasienergies. 
 The emitted phonons can be blue shifted $I_{\rm{b(lue)}}$, unshifted or red shifted $I_{\rm{r(ed)}}$, respectively. Clearly, the unshifted transitions do no change the system state, but the shifted transitions can be used to control it.
 
  Our main findings are summarized in Fig.~\ref{fig:overview} (c) and (d), where we depict the probability of the Floquet state with the lowest quasienergy $p_0$ and the difference of emitted phonons $I_{\rm{b}}-I_{\rm{r}}$ as a function of driving amplitude $h_{z,1}$ and a system-environment coupling angle $\theta$. Remarkably, we observe jumps in both observables at specific  values of the driving amplitude, marked with arrows in Fig.~\ref{fig:overview} (c) and (d). As we will explain in detail, they appear at parameters, at which CDT takes place.

\textbf{Model system.}
We consider a generic two-level system, which describes, e.g., the low-energy physics of a double-well potential or a superconducting qubit. It is externally driven and coupled to a thermal environment as sketched in Fig.~\ref{fig:overview}(a). The Hamiltonian reads
\begin{equation}
 H(t) = \frac{h_{\rm x} }{2}\sigma_{x} + \frac{h_{\rm z}(t)}{2}\sigma_{\rm z} +  \hat \sigma_{\theta} \sum_{k}V_k\left(b_{k} + b_{k}^\dagger\right) + H_{\rm B},
\end{equation}
where $\sigma_{\alpha}$ with $\alpha= \left\lbrace \rm x,y,z \right\rbrace$ denote the Pauli matrices, $h_{\rm x}$ denotes the tunneling amplitude and $h_{\rm z}(t)= h_{\rm z,0} + h_{\rm z,1} \cos (\Omega t)$ is the time-dependent  energy splitting, where $h_{\rm z,0}$ is the offset, $h_{\rm z,1}$  is the driving  amplitude and $\Omega$ is the driving frequency. The bath is quadratic in bosonic operators $b_k$ and is coupled via the  system operator  $\hat \sigma_{\theta} =  \sin\theta \sigma_{\rm x} +  \cos\theta \sigma_{\rm z} $ with strengths $V_k$. Depending on the coupling angle $ \theta$, it is known that undriven systems can give rise to diverse physical behavior~\cite{Duan,CastroNeto2003,Guo2012,Kohler2013,Bruognolo2014}.

Floquet theory describes the dynamics of periodically-driven systems~\cite{Shirley1965,Grifoni1998}. Due to the periodic system  dynamics with frequency $ \Omega=2\pi/\tau $, there are characteristic states  of the system which fulfill $\left|\Phi_{n,\lambda}(t) \right> =e^{-i \epsilon_{n,\lambda} t} \left|\varphi_{n,\lambda}(t) \right>$, with quasienergy $\epsilon_{n,\lambda}$ and periodic Floquet state $  \left|\varphi_{n,\lambda}(t) \right> =  \left|\varphi_{n,\lambda}(t+\tau) \right>$. These states are the analogue to the eigenstates in time-independent systems. Importantly, the Floquet states $\lambda$ are not uniquely defined due to the Brillouin zone index $n$: a state with index $n$ can be related to the $n=0$ state by $\epsilon_{n,\lambda} = \epsilon_{0,\lambda} + n \Omega  $ and $  \left|\varphi_{n,\lambda}(t) \right>=  e^{-i n \Omega t} \left|\varphi_{0,\lambda}(t) \right>$. The stroboscopic Floquet states are the eigenstates of the time-evolution operator after one period $\hat U_{\rm s}(\tau) \left|\varphi_{n,\lambda}(0) \right> = e^{-i \epsilon_{n,\lambda} \tau }\left|\varphi_{n,\lambda}(0) \right> $. 

\begin{figure}[t]
	\includegraphics[width=1\linewidth]{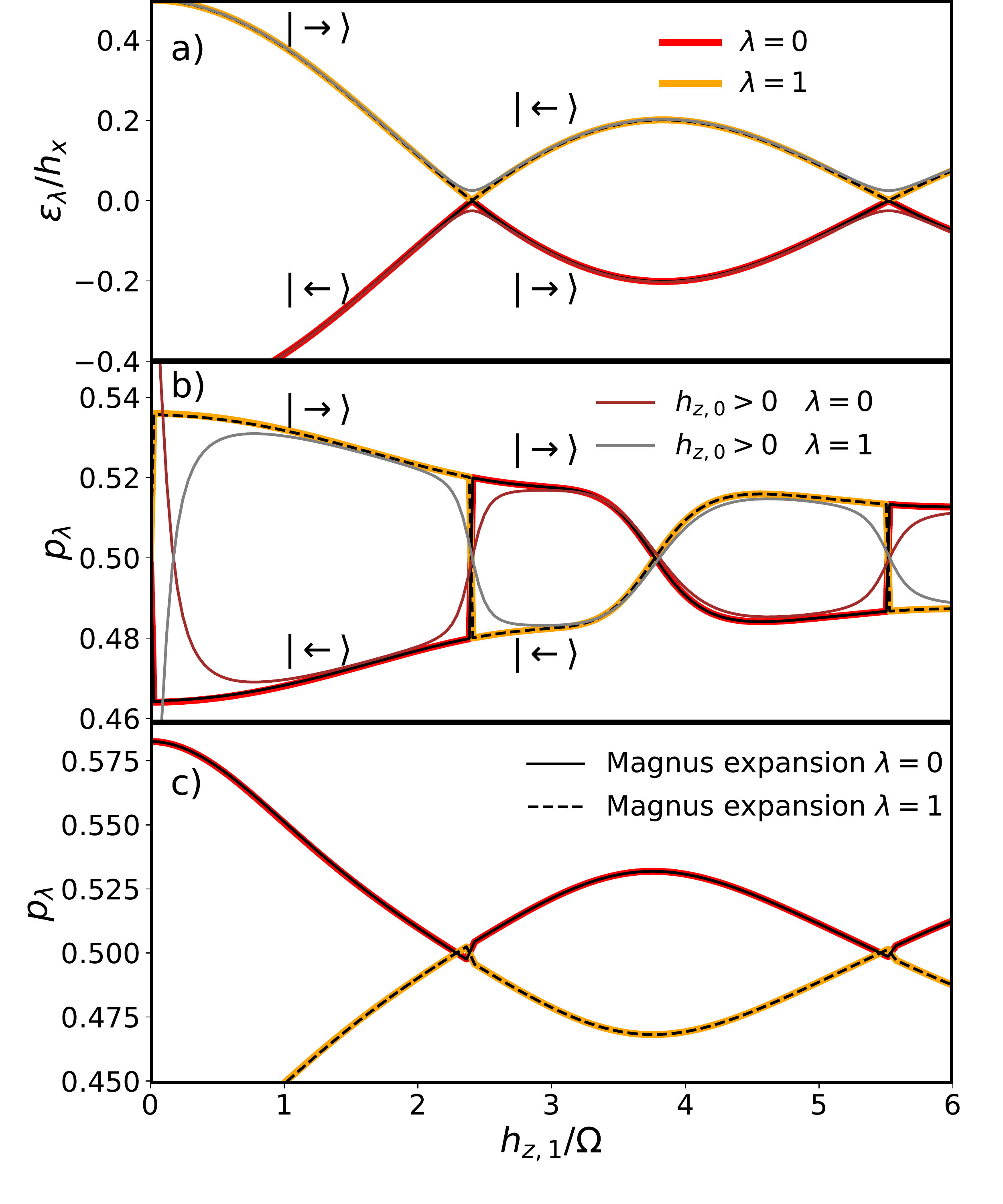}
	\caption{ (a) The quasienergy spectrum as a function of the driving amplitude $h_{\rm z,1}$.   (b) and (c) depict slices of Fig.~\ref{fig:overview}(c) showing both stationary Floquet state probabilities $p_\lambda$ for $\theta=\pi/2$  ( $\hat \sigma_{\pi/2}= \sigma_{\rm x}$)  and $\theta=\pi/4$ ( $\hat \sigma_{\pi/4}= \frac{1}{\sqrt{2}}\sigma_{\rm x} + \frac{1}{\sqrt{2}} \sigma_{\rm z}$), respectively.  The Floquet states are approximately given by the eigenstates of $\sigma_{\rm x}$, which we denote by $\left|-1 \right>_{\rm x}$,$\left|+1 \right>_{\rm x}$ corresponding to eigenvalues $-1$,$1$, respectively. 
	}
	\label{fig:steadyState}
\end{figure}

In Fig.~\ref{fig:steadyState}(a), we depict the numerically calculated quasienergies $\epsilon_{\lambda} =  \epsilon_{0,\lambda}$ as function of  $h_{\rm z,1}/ \Omega $. Here, the index $\lambda$ can take the values $\lambda=0,1$ and we consider  $n=0$. The stroboscopic dynamics follows the effective Hamiltonian $H_{\rm eff} = \frac{h_{\rm z,0}} 2 \sigma_{\rm z} + \frac{h_{\rm x}} 2 \mathcal J_0 (h_{\rm z,1}/ \Omega) \sigma_{\rm x} + \mathcal O(\frac{1}{\Omega}) $.  For $h_{\rm z,0}=0$, we find $\epsilon_{\lambda} = \pm h_{\rm x} \mathcal J_{0} (h_{\rm z,1}/ \Omega)/2$, so that there are degeneracies at the roots of the Bessel function  $\mathcal J_{0} (h_{\rm z,1}/ \Omega)=0$.  This is the celebrated CDT effect, as the dynamics at these parameters is frozen. The stroboscopic Floquet states read $  \left|\varphi_{\lambda}(0) \right>  \approx \left|\text{sign}\left[ \mathcal J_0  (h_{\rm z,1}/ \Omega)\right](-1)^{\lambda} \right>_{\rm x}$. Accordingly, there is a non analytic crossover of the Floquet state, e.g. $\left|\varphi_{0}(0) \right> = \left|-1 \right>_{\rm x}$ to $\left|\varphi_{0}(0) \right>= \left|+1 \right>_{\rm x}$, at the roots of the Bessel function. As we explain, the crossover of the Floquet states  is the origin of the probability jumps observed in Fig.~\ref{fig:overview}.

\textbf{Rate equations.}
An important point to realize is that  though the states $  \left|\varphi_{n,\lambda}(t) \right>$ of the spin system  are equivalent for different $n$ in a closed system, $n$ becomes physically relevant when  the system is coupled to a thermal bath  $ H_{\rm B}$. In this situation, the bath can trigger transitions between different Brillouin zones $n,n'$. In Fig.~\ref{fig:overview}(b) we illustrate transitions associated with $\Delta n= n -n'=-1$. Using the secular Floquet-Redfield formalism~\cite{Grifoni1998,Carreno2017}, one can derive the rate equations
\begin{align}
\frac d {dt } p_0 &= - \sum_{\Delta n} A_{1\leftarrow 0 }^{(\Delta n)} p_0 +\sum_{\Delta n} A_{0\leftarrow 1 }^{\Delta n} p_1 \nonumber , \\
\frac d {dt } p_1 &=  + \sum_{(\Delta n)} A_{1\leftarrow 0 }^{(\Delta n)} p_0 -\sum_{\Delta n} A_{0\leftarrow 1 }^{(\Delta n)} p_1 ,
\label{eq:rateEquation}
\end{align}
where $p_\lambda$ denotes the probability to be in Floquet state $\lambda$ and $A_{\lambda\leftarrow\mu}^{(n)}$ is the transition probability between Floquet states 
\begin{align}
 A_{\lambda\leftarrow\mu}^{(n)} &= \Gamma (\Delta_{\lambda\mu}^{n})  \left[n_B\left( \Delta_{\lambda\mu}^{n} \right) +1 \right] \cdot \left| a_{\lambda\leftarrow \mu}^{n} \right|^2 ,\nonumber \\
 a_{\lambda\leftarrow \mu}^{(n)} &= \frac 1 \tau \int_{0}^{\tau}  \left<\varphi_{\lambda}(0) \right|  \hat  \sigma_{\theta}   (t) \left|\varphi_{\mu}(0) \right> e^{-i n \Omega t} dt .
 \label{eq:rates}
\end{align}
Here, $n_{\rm B}(\omega)$ denotes the Bose distribution and $\Gamma (\omega) = \sum_{k}V_k^2 \delta(\omega-\omega_k)= \gamma \omega/(\omega^2+\omega_c^2)$ (coupling strength $\gamma$, cut-off frequency $\omega_c$) denotes the coupling density, which we define for negative frequencies by $\Gamma (\omega)=-\Gamma (-\omega)$, and $\Delta_{\lambda\mu}^{n} = \epsilon_{\mu}-\epsilon_{\lambda}- n\Omega$. The time-dependent operator reads $ \hat \sigma_{\theta}  (t)=    e^{i\hat \Lambda (t)}   \hat \sigma_{\theta}      e^{- i \hat \Lambda (t)}$, where $\hat \Lambda(t)$ is defined by
\begin{align}
	\left| \varphi_\lambda (t)\right>  =  e^{-i\hat \Lambda (t)} \left| \varphi_\lambda (0)\right>,
\end{align}
at which the operator $ e^{-i \hat \Lambda (t)} $ propagates the Floquet states.

It is easy to show that the  coefficients fulfill $ a_{\lambda\leftarrow \mu}^{(n)} =\left( a_{\mu\leftarrow \lambda}^{(-n)}\right)^*$. As  a consequence, the related  rates obey the detailed balance condition $ A_{\lambda\leftarrow\mu}^{(n)} = A_{\mu\leftarrow\lambda}^{(-n)} e^{\Delta_{\lambda\mu}^{(n)}/ T}$, where $T$ is the temperature of the environment. Yet, in general 
$\left|a_{\mu\leftarrow  \lambda}^{(n)} \right|  \neq \left|a_{\lambda\leftarrow \mu}^{(n)} \right| $, which gives rise to a break down of the detailed balance relation in the stationary state, thus $ p_0 /p_1 \neq e^{-\Delta / T}$ with $\Delta = \epsilon_1 - \epsilon_0$.

\begin{figure}[t]
\includegraphics[width=1\linewidth]{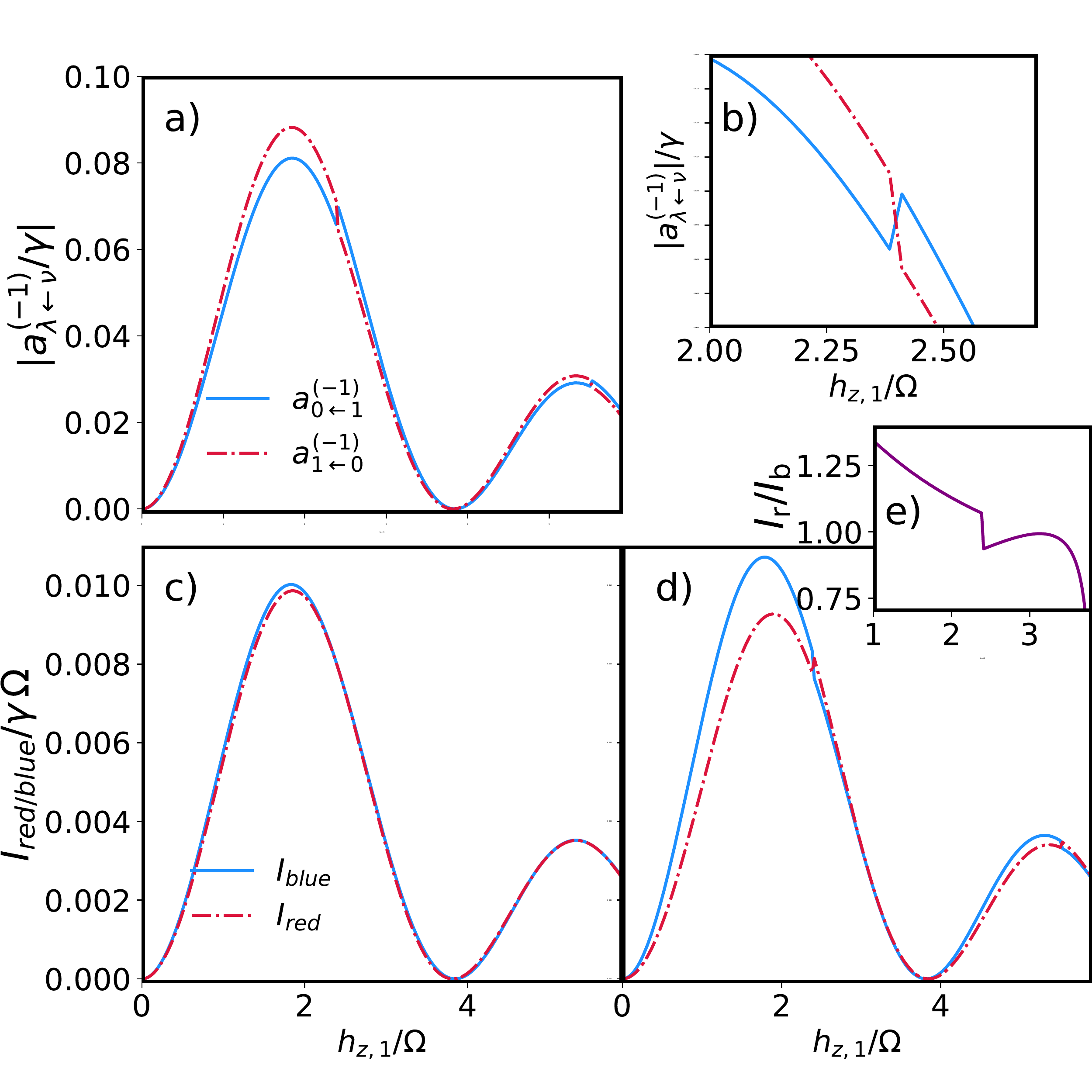}
\caption{ (a) The transition coefficients  for $\hat\sigma_{\pi/2}$ with parameters equal to Fig.~\ref{fig:overview}(c). The coefficients for  $\hat\sigma_{\pi/4}$ are almost equal. (b) Magnification of the jump discontinuity in (a). (c) and (d) depict corresponding intensities of blue and red shifted emitted phonons for $\theta =\pi/2 $ and $\theta =\pi/4$ coupling, respectively. The ratio of the red and blue shifted emission is depicted in (e), their difference can be found in Fig.~\ref{fig:overview}(d).
}
\label{fig:transistions}
\end{figure}

\textbf{Stationary state.}
Figures~\ref{fig:steadyState}(b) and (c) depict the stationary state probabilities for the coupling $\hat \sigma_\theta$  with $\theta = \pi/2,\pi/4$, respectively. For $\theta = 0,\pi$, the system approaches a Floquet-Gibbs state according to Ref.~\cite{Shirai2016}.  In (b), we find   a probability inversion for small $h_{\rm z,1}/\Omega$ and  probability jumps at the roots of the Bessel function $\mathcal J_{0}(h_{\rm z,1}/\Omega)=0$. These effects can be explained by analyzing the rates in Eq.~\eqref{eq:rates}.  Due to a generalized parity symmetry,  $A_{\lambda\leftarrow\mu}^{(0)} $ vanishes exactly~\cite{parityNote}. Consequently, the rate equations are dominated by the transitions $\Delta n =-1$, as  $n_{\rm B}(|\Delta^{n>0}_{\lambda \mu}|)\ll 1 $  for  $\Delta n>0$  due to  a large energy difference $\Delta_{\lambda,\mu}^{n}\approx - n \Omega$, i.e.,  the transitions marked by the red and blue arrows in Fig.~\ref{fig:overview}(b) (green transitions do not change the state). The corresponding coefficients $\left|a_{\mu\leftarrow  \lambda}^{(-1)} \right|^2$ are depicted in Fig.~\ref{fig:transistions}(a), where we observe that $\left|a_{1 \leftarrow  0}^{(-1)} \right|^2>\left|a_{0\leftarrow  1}^{(-1)} \right|^2$ for, e.g., $h_{z,1}/\Omega<z_0$, with $z_0$ denoting the first root of the zeroth-order Bessel function. As $n_{\rm B}(\Delta^{n}_{\lambda \mu})\ll 1 $ and $\Gamma(\Delta^{(-1)}_{10} )\approx \Gamma(\Delta^{(-1)}_{01} ) $, we find from Eq.~\eqref{eq:rateEquation} that $p_1/p_0 \approx \left|a_{1 \leftarrow  0}^{(-1)} \right|^2/ \left|a_{0\leftarrow  1}^{(-1)} \right|^2$ which explains the probability inversion. 
The rates  explain the jump in the probability distribution. In Fig.~\ref{fig:transistions}(a)  we observe jumps   at the CDT positions. The jump is magnified in Fig.~\ref{fig:transistions}(b). The non-continuous behavior becomes more clear when considering Eq.~\eqref{eq:rates}. At the CDT, the Floquet states switch, thus $\left|\varphi_{\lambda}(0) \right> \leftrightarrow \left|\varphi_{\mu}(0) \right> $, which gives rise to the non-analytic behavior. 

A similar reasoning can  be applied to the  $\hat \sigma_{\pi/4}$ coupling depicted in Fig.~\ref{fig:steadyState}(c). Away from the CDT, the probability distribution mainly corresponds to the Floquet-Gibbs state. This appears as the coupling $\hat \sigma_{ \pi/4}$ has a $\sigma_z$ contribution so that the rates $A_{\lambda\leftarrow\mu}^{(0)} \gg A_{\lambda\leftarrow\mu}^{(-1)} $ result mainly in   a Gibbs state. However, the coefficients $a_{\lambda\leftarrow  \mu}^{(-1)}$ are almost equal to those of the $\hat \sigma_{ \pi/2}$  case  and thus give rise to a probability jump, yet, to  a very small extend.

Importantly, although there is a jump discontinuity, the density matrix remains continuous as a function of $h_{\rm z,1}$. In the Floquet-Redfield formalism, the system density matrix reads $\rho_{\rm s}(t)= \sum_{\lambda} p_{\lambda}(t) \left| \varphi_{\lambda}(t)\right>\left< \varphi_{\lambda }(t) \right| $. As there is a simultaneous switch of $p_{\lambda}$ and $\left| \varphi_{\lambda}(t)\right>$, the density matrix remains continuous. Yet, the probability jump  does not depend on how  the states $\lambda =0,1$ are labeled, as the labeling can be uniquely defined.
Let us consider the system with $h_{\rm z,0}\neq 0$. A corresponding quasienergy spectrum is depicted in Fig.~\ref{fig:steadyState}(a) with thin lines. For finite but small  $h_{\rm z,0} $ the gap closing is released as depicted in Fig.~\ref{fig:steadyState}(a). Accordingly, the states depend smoothly on $h_{\rm z,1}$, so that the $p_\lambda$ are also uniquely defined as can be observed in Fig.~\ref{fig:steadyState}(b). The ordering of the states $\lambda=0,1$ can be thus uniquely defined in terms of the limit $h_{\rm z,0}\rightarrow 0$.  

The interplay of $\theta$ and $h_{\rm z,1}$ can be fully analyzed in Fig.~\ref{fig:overview}(c), where the stationary state strongly depends on $\theta$, though, the system is only weakly coupled to the environment. In particular, for $\theta\approx 0.5\pi$ and small $h_{\rm z,1}$ we find inversion, $p_0 < 0.5 $, thus, there is a strong deviation from the Floquet-Gibbs state, which can be found for $\theta=0$.


\textbf{Phonon emission.}
Due to the continuous stationary state, the jump cannot be observed in system observables. However, the nonanalytic behavior can be observed in the emission. Every transition $A_{\lambda\leftarrow\mu}^{(-1)}  $ is related to the emission of phonons with either energy $\Omega$,$\Omega\pm \Delta$. The corresponding intensities  $I_{\rm b,r} =I(\Omega\pm\Delta)$ are given by 
$
I_{\rm b} =  \left(\Omega + \epsilon_1 -\epsilon_0 \right) A_{0\leftarrow 1}^{(-1)} p_{1},  
 I_{\rm r} = \left(\Omega - \epsilon_1 +\epsilon_0 \right)  A_{1\leftarrow 0}^{(-1)} p_{0}.
$
We depict the blue and red shifted intensities in Fig.~\ref{fig:transistions} (c) and (d). For $\hat \sigma_{ \pi/ 4}$, we observe that  the two intensities are almost  equal.  As the stationary state is governed by the rates $A_{\lambda\leftarrow \nu}^{(-1)}$, we find $p_{0}\propto A_{0\leftarrow 1}^{(-1)} $ and $p_{1}\propto A_{1\leftarrow 0}^{(-1)} $, so that $I_{\rm r}\approx I_{\rm b}$ for high frequency $\Omega\gg  \epsilon_1 -\epsilon_0$. 

For $\hat \sigma_{ 0 } = \sigma_{z}$, at which the system approaches a Floquet-Gibbs state, it is known that the rates $  A_{\mu\leftarrow \nu}^{n\neq 0} \approx0$~\cite{Shirai2015}. Consequently, here both  $I_{\rm r/b}$ vanish.
Considering the difference $ I_{\rm b }- I_{\rm r}$  in Fig.~\ref{fig:overview}(d), we consequently find  that the difference is  smooth in $h_{\rm z,1}$ for both limiting cases $\theta=0,\pi/2$. However, in between there is a significant jump, which is strongest for about $\theta\approx 0.3 \pi$.

Let us consider the $\hat \sigma_{\pi/4}$ coupling. Due to the $\sigma_z$ coupling component, the $A_{\lambda\leftarrow \nu}^{( 0) }$ rates are dominant, which leads to an (almost) thermalization of the system with its environment. However, the rates $A_{\lambda\leftarrow \nu}^{(-1)}$ still exhibit a jump at the CDT, so that we find jumps in  $I_{\rm r }$ and $ I_{\rm b}$, as can be observed in Fig.~\ref{fig:transistions}(d).

%

\textbf{Magnus expansion.}
To find an appropriate approximation of the numerical results, we employ a rotating wave approximation  defined by a unitary transformation~\cite{Gong2009} 
$U_{\rm rot}(t) = \exp \left[- i  \sigma_{z}   \theta(t)  \right]  $
with $\theta (t) = \frac{h_{\rm z,1} }{\Omega} \sin(\Omega t  )  $.
The resulting Hamiltonian  $H_r (t)= U_{\rm rot}^\dagger \hat H(t) U_{\rm rot} $ also exhibits  $\tau$ periodicity. The Floquet state propagator can be written as
\begin{align}
 e^{-i \hat \Lambda (t)} &= \hat U_{\rm rot}(t)  e^{-i \hat \Lambda_{\rm r} (t)}   
 					\label{eq:lambdaInRotFrame}
 					\approx \hat U_{\rm rot}(t) \left[ 1-i \hat \Lambda_{\rm r} (t) \right] 
%
\end{align}
with $e^{-i \hat \Lambda_{r} (t)}$ being the Floquet state propator in the rotating frame. The explicit form of $\hat \Lambda_{r}$ can be calculated using a standard  high frequency expansion \cite{Mananga2011}.  However, using  Eq.~\eqref{eq:lambdaInRotFrame} without approximation, it is not possible to evaluate the integral in Eq.~\eqref{eq:rates}. As  $\hat \Lambda_{r} \propto \mathcal O(\frac{1}{\Omega})$ is  small in the high frequency limit, the expansion in Eq.~\eqref{eq:lambdaInRotFrame} is justified. For consistency, we also expand the Hamiltonian up to the same order, thus $H_{\rm eff}=H_{\rm eff}^{(0)} + \frac 1 \Omega H_{\rm eff}^{(1)} + \mathcal O \left( \frac{1}{\Omega^2} \right)$. Evaluating Eq.~\eqref{eq:rates}, we obtain
\begin{align}
 a_{\mu,\lambda}^{(n)}	=\mathcal S^{n}_{\rm x} \left< \sigma_{\rm x}  \right>_{\mu\lambda} +i\cdot \mathcal S^{n}_{\rm y}\left< \sigma_{\rm y}  \right>_{\mu\lambda}+\mathcal S^{n}_{\rm z} \left< \sigma_{\rm z}  \right>_{\mu\lambda},
 \label{eq:analyticCoefficients}
\end{align}
where we have defined $\left< \sigma_{\alpha}  \right>_{\mu\lambda}= \left< u_{\mu}(0)\right| \sigma_{\alpha}   \left| u_{\lambda}(0)\right>$ and

\begin{align}
	\mathcal S^{n}_{\rm x} &= g_{\rm x} \delta_{n\,\text{mod}\,2,0} \mathcal J_{n} \left(  \frac{ h_{\rm z,1} }{\Omega} \right)
	-2  g_{\rm z} \delta_{n,0}
	 \sum _{k=1}^{\infty} l_{2k-1},\label{eq:analyticRates}\\
	  &+g_{\rm z} \delta_{n \text{mod} 2, 1} l_{|n|} \nonumber ,\\
	\mathcal S^{n}_{\rm y} &=  g_{\rm x}\delta_{n\,\text{mod}\,2,1} \mathcal J_{n} \left(  \frac{ h_{\rm z,1} }{\Omega}\right)-g_{\rm z} \delta_{n \text{mod} 2, 0} l_{|n|} \nonumber ,\\
	\mathcal S^{n}_{\rm z} &=g_z \delta_{0,n} -g_{\rm z}  \mathcal J_{-n-2k} \left(   \frac{ h_{\rm z,1} }{\Omega} \right)   \delta_{n \,\text{mod}\, 2, 0}\sum _{k=1}^{\infty} l_{2k-1} \nonumber , \\
	&+ g_{\rm x} \frac 12   \delta_{n \,\text{mod}\, 2, 1} \alpha_{\rm x}^{(n)} \nonumber ,\\
	\alpha_{\rm x}^{(n)}  &= \sum _{m=1}^{\infty} l_{m} \left[
	\mathcal J_{-n+m} \left(   \frac{ h_{\rm z,1}}{\Omega} \right) \nonumber 
	-(-1)^{m}  \mathcal J_{-n-m} \left(   \frac{ h_{\rm z,1} }{\Omega} \right)   
	\right],
	\nonumber
\end{align}
with $g_{\rm x}=\sin\theta$, $g_{\rm z}=\cos\theta$, $l_m=\frac{h_x }{m \Omega}\mathcal J_{m}\left( h_z/ \Omega\right)$ for $m>0$ and $l_m=0 $ for $m=0$.  Due to the factor $h_{\rm x} / \Omega$ the coefficients $l_n$ can be considered to be small.

 The transition coefficients $a_{\mu,\lambda}^{(n)}$ in Eq.~\eqref{eq:analyticRates} are evaluated  in Tab.~\ref{tab:transitionCoef}.  For $n=0$, we find a hermitian  structure for  the eigenstates  $\left|\varphi_{\lambda}(0) \right> \approx \left|-1 \right>_{\rm x},\left|+1 \right>_{\rm x} $, which are mainly determined by $H_{\rm eff}^{(0)}\propto \sigma_{\rm x}$. For $n=-1$, the transition coefficients are dominated by $\mathcal S_{\rm y}^{(-1)}\propto \mathcal J_{-1}(h_{z,1}/\Omega)$, which explains the oscillations in Fig.~\ref{fig:transistions}(a). Importantly, they   do not exhibit a hermitian structure. This leads to a breakdown of the detailed balance relation, and  gives rise to the jump in the probability distribution  in Fig.~\ref{fig:steadyState}(b) and (c). This appears as  $\left|\varphi_{0}(0) \right>$ switches  from $\left|-1 \right>_{\rm x}$ to $ \left|+1 \right>_{\rm x} $ and, simultaneously,
 $\left|\varphi_{1}(0) \right>$ switches from  $\left|+1 \right>_{\rm x}$ to $\left|-1 \right>_{\rm x}$, causing a jump of $a_{\mu,\lambda}^{(-1)}$.

We can use Eq.~\eqref{eq:analyticCoefficients} to understand the intensity jump in the high-frequency regime. Assuming that $p_\lambda\approx 0.5$ as  in Fig.~\ref{fig:steadyState}(b), we find for $h_{\rm z,1}/\Omega\approx z_0$
\begin{align}
 \frac{I_{\rm red}}{I_{\rm blue}} \approx \frac{\left(\mathcal J_{-1} \left(  z_0 \right) -\frac {h_{\rm x} }  {\Omega}\alpha_{\rm x}^{(-1)}(z_0)  \right)^2 } {\left(\mathcal J_{-1} \left(  z_0 \right) +\frac {h_x}  {\Omega}\alpha_{\rm x}^{(-1)}(z_0)  \right)^2 } \approx 1\pm 2 \frac {h_x}  {\Omega}\alpha_{\rm x}^{(-1)}(z_0)
 \nonumber
\end{align}
with a constant $\alpha_{\rm x}^{(-1) }(z_0)$, and '$\pm$' for $h_z/\Omega \lessgtr z_0$. Thus, interestingly, this ratio (Fig.~\ref{fig:transistions}(e)) close to the CDT  and consequently  the jump magnitude $4 \frac {h_x}  {\Omega}\alpha_{\rm x}^{(-1)}(z_0)$ scales as $1/\Omega$. 

\begin{table}[t]
\caption{Coupling coefficients $a_{\mu,\lambda}^{(n)}  $ of all possible transitions between different Floquet states $\left|\varphi_\lambda(0) \right>$.}
\small
\begin{tabular}{c|cc}
$n=0$             &  $\left|-1 \right>_{\rm x} $ & $\left|+1 \right>_{\rm x} $ \\
\hline
$\left|-1 \right>_{\rm x} $  & $ \mathcal S_{\rm x}^{(0)} $ &$ \mathcal  S_{\rm z}^{(0)} $ \\
$\left|+1 \right>_{\rm x} $ & $\mathcal S_{\rm z}^{(0)} $ & $-\mathcal  S_{\rm x}^{(0)} $
\end{tabular}
\hspace{0.2cm}
\begin{tabular}{c|cc}
$-1$            &  $\left|-1 \right>_{\rm x} $ & $\left|+1 \right>_{\rm x} $ \\
\hline
$\left|-1 \right>_{\rm x} $  & $ \mathcal S_x^{(-1)} $ &$ \mathcal S_{\rm y}^{(-1)} +\mathcal S_z^{(-1)}$ \\
$\left|+1 \right>_{\rm x} $ & $\mathcal S_y^{(-1)} -\mathcal S_{\rm z}^{(-1)} $ & $-\mathcal S_x^{(-1)} $
\end{tabular}
\label{tab:transitionCoef}

\end{table}


\textbf{Discussion.}
The CDT in a driven dissipative system gives rise to surprising effects. Besides the well-investigated freezing of the  internal system dynamics at the CDT,  the presence of the thermal environment can give rise to counter intuitive jumps in the probability distribution of the Floquet state. Yet, as there is a simultaneous switch of Floquet states and probability, the reduced density matrix remains continuous while crossing the CDT. Consequently, the probability jumps can not be observed in system observables. 
Moreover, the jump behavior has a drastic consequence on the Mollow triplet, such that the blue and red shifted intensities both exhibit a  discontinuity at the CDT. This can be  directly measured by  phonon  spectroscopy. The ratio of both shifted intensities exhibits a jump of about $10\%$.

The underlying physical reason for these discontinuities is a switch of the Floquet states at the CDT. This causes a jump in the system-bath coupling coefficients. Consequently, the effect does not depend on the details of the thermal bath. Yet, the system-bath coupling operator is important. For a pure $\hat \sigma_0= \sigma_z$ coupling as investigated in Ref.~\cite{Shirai2015}, the stationary  density matrix of the system recovers an effective Floquet-Gibbs states. For $\hat \sigma_{\pi/2}= \sigma_x$, we find extreme deviations from the Floquet-Gibbs state with occupation inversion even for small driving amplitude as observed in Fig.~\ref{fig:steadyState}(b). With this coupling, the probability jump at the CDT turns out to be most significant. However, there is no signature in the emitted  phonons.  For $\hat \sigma_{\pi/4}$ coupling, though the probability jumps are very small, there is a clear jump discontinuity in the blue and red shifted intensity.

The noncontinuous behaviour is not a consequence of the high-frequency regime, which we considered here to explain the numerical results with analytical calculations. The CDT appears due to an exact degeneracy of the quasienergyies which is persistant even for very low driving-frequencies~\cite{Grifoni1998}. Consequently, the jump behavior will remain when lowering the driving frequency. Our findings are not restricted to the dissipative two-level system. Similar probability jumps could be also observed in dissipative driven Lipikin-Meshkov-Glick model, which gives  rise to many-body CDT~\cite{Gong2009}. Furthermore, these findings will be important for electronic transport~\cite{Sanchez2008,Brandes2004,Sanchez2007}. It will be interesting to explore the fate of the jumps for stronger environmental coupling using methods such as~\cite{Restrepo2016,Magazzu2018,Xu2016,Duan2017,Lee2012}

Finally, the analytical treatment based on the Magnus expansion extended with our approximation Eq.~\eqref{eq:lambdaInRotFrame} is the key tool to understand the discontinuities. The analytical treatment can generalized  to other configurations of the driving which could give rise to even more flexible mechanisms of the stationary state, and other exotic stationary states in driven systems.

\textbf{Acknowledgments.}
G. Engelhardt gratefully acknowledges financial support
from the China Postdoc Science Foundation (Grant No.: 2018M640054) , J. Cao  acknowledges support from the NSF (Grant No.: CHE 1800301 and CHE 1836913). Both acknowledge the Natural Science Foundation of China (under Grant No.:U1530401).
G. Platero acknowledges MINECO (Grant No.: MAT2017-86717-P).  The authors thank Hui Dong and Tilen \v Cade\v z for inspiring discussions

\bibliography{mybibliography}

\begin{thebibliography}{42}%
\makeatletter
\providecommand \@ifxundefined [1]{%
 \@ifx{#1\undefined}
}%
\providecommand \@ifnum [1]{%
 \ifnum #1\expandafter \@firstoftwo
 \else \expandafter \@secondoftwo
 \fi
}%
\providecommand \@ifx [1]{%
 \ifx #1\expandafter \@firstoftwo
 \else \expandafter \@secondoftwo
 \fi
}%
\providecommand \natexlab [1]{#1}%
\providecommand \enquote  [1]{``#1''}%
\providecommand \bibnamefont  [1]{#1}%
\providecommand \bibfnamefont [1]{#1}%
\providecommand \citenamefont [1]{#1}%
\providecommand \href@noop [0]{\@secondoftwo}%
\providecommand \href [0]{\begingroup \@sanitize@url \@href}%
\providecommand \@href[1]{\@@startlink{#1}\@@href}%
\providecommand \@@href[1]{\endgroup#1\@@endlink}%
\providecommand \@sanitize@url [0]{\catcode `\\12\catcode `\$12\catcode
  `\&12\catcode `\#12\catcode `\^12\catcode `\_12\catcode `\%12\relax}%
\providecommand \@@startlink[1]{}%
\providecommand \@@endlink[0]{}%
\providecommand \url  [0]{\begingroup\@sanitize@url \@url }%
\providecommand \@url [1]{\endgroup\@href {#1}{\urlprefix }}%
\providecommand \urlprefix  [0]{URL }%
\providecommand \Eprint [0]{\href }%
\providecommand \doibase [0]{http://dx.doi.org/}%
\providecommand \selectlanguage [0]{\@gobble}%
\providecommand \bibinfo  [0]{\@secondoftwo}%
\providecommand \bibfield  [0]{\@secondoftwo}%
\providecommand \translation [1]{[#1]}%
\providecommand \BibitemOpen [0]{}%
\providecommand \bibitemStop [0]{}%
\providecommand \bibitemNoStop [0]{.\EOS\space}%
\providecommand \EOS [0]{\spacefactor3000\relax}%
\providecommand \BibitemShut  [1]{\csname bibitem#1\endcsname}%
\let\auto@bib@innerbib\@empty
\bibitem [{\citenamefont {Aidelsburger}\ \emph {et~al.}(2013)\citenamefont
  {Aidelsburger}, \citenamefont {Atala}, \citenamefont {Lohse}, \citenamefont
  {Barreiro}, \citenamefont {Paredes},\ and\ \citenamefont
  {Bloch}}]{Aidelsburger2013}%
  \BibitemOpen
  \bibfield  {author} {\bibinfo {author} {\bibfnamefont {M.}~\bibnamefont
  {Aidelsburger}}, \bibinfo {author} {\bibfnamefont {M.}~\bibnamefont {Atala}},
  \bibinfo {author} {\bibfnamefont {M.}~\bibnamefont {Lohse}}, \bibinfo
  {author} {\bibfnamefont {J.~T.}\ \bibnamefont {Barreiro}}, \bibinfo {author}
  {\bibfnamefont {B.}~\bibnamefont {Paredes}}, \ and\ \bibinfo {author}
  {\bibfnamefont {I.}~\bibnamefont {Bloch}},\ }\bibfield  {title} {\enquote
  {\bibinfo {title} {{Realization of the Hofstadter Hamiltonian with ultracold
  atoms in optical lattices}},}\ }\href {\doibase
  10.1103/PhysRevLett.111.185301} {\bibfield  {journal} {\bibinfo  {journal}
  {Phys. Rev. Lett.}\ }\textbf {\bibinfo {volume} {111}},\ \bibinfo {pages}
  {185301} (\bibinfo {year} {2013})}\BibitemShut {NoStop}%
\bibitem [{\citenamefont {Jotzu}\ \emph {et~al.}(2014)\citenamefont {Jotzu},
  \citenamefont {Messer}, \citenamefont {Desbuquois}, \citenamefont {Lebrat},
  \citenamefont {Uehlinger}, \citenamefont {Greif},\ and\ \citenamefont
  {Esslinger}}]{Jotzu2014}%
  \BibitemOpen
  \bibfield  {author} {\bibinfo {author} {\bibfnamefont {G.}~\bibnamefont
  {Jotzu}}, \bibinfo {author} {\bibfnamefont {M.}~\bibnamefont {Messer}},
  \bibinfo {author} {\bibfnamefont {R.}~\bibnamefont {Desbuquois}}, \bibinfo
  {author} {\bibfnamefont {M.}~\bibnamefont {Lebrat}}, \bibinfo {author}
  {\bibfnamefont {T.}~\bibnamefont {Uehlinger}}, \bibinfo {author}
  {\bibfnamefont {D.}~\bibnamefont {Greif}}, \ and\ \bibinfo {author}
  {\bibfnamefont {T.}~\bibnamefont {Esslinger}},\ }\bibfield  {title} {\enquote
  {\bibinfo {title} {Experimental realization of the topological {Haldane}
  model with ultracold fermions},}\ }\href@noop {} {\bibfield  {journal}
  {\bibinfo  {journal} {Nature (London)}\ }\textbf {\bibinfo {volume} {515}},\
  \bibinfo {pages} {237--240} (\bibinfo {year} {2014})}\BibitemShut {NoStop}%
\bibitem [{\citenamefont {Benito}\ \emph {et~al.}(2014)\citenamefont {Benito},
  \citenamefont {G\'omez-Le\'on}, \citenamefont {Bastidas}, \citenamefont
  {Brandes},\ and\ \citenamefont {Platero}}]{Benito2014}%
  \BibitemOpen
  \bibfield  {author} {\bibinfo {author} {\bibfnamefont {M.}~\bibnamefont
  {Benito}}, \bibinfo {author} {\bibfnamefont {A.}~\bibnamefont
  {G\'omez-Le\'on}}, \bibinfo {author} {\bibfnamefont {V.~M.}\ \bibnamefont
  {Bastidas}}, \bibinfo {author} {\bibfnamefont {T.}~\bibnamefont {Brandes}}, \
  and\ \bibinfo {author} {\bibfnamefont {G.}~\bibnamefont {Platero}},\
  }\bibfield  {title} {\enquote {\bibinfo {title} {Floquet engineering of
  long-range $p$-wave superconductivity},}\ }\href {\doibase
  10.1103/PhysRevB.90.205127} {\bibfield  {journal} {\bibinfo  {journal} {Phys.
  Rev. B}\ }\textbf {\bibinfo {volume} {90}},\ \bibinfo {pages} {205127}
  (\bibinfo {year} {2014})}\BibitemShut {NoStop}%
\bibitem [{\citenamefont {Engelhardt}\ \emph {et~al.}(2017)\citenamefont
  {Engelhardt}, \citenamefont {Benito}, \citenamefont {Platero},\ and\
  \citenamefont {Brandes}}]{Engelhardt2017a}%
  \BibitemOpen
  \bibfield  {author} {\bibinfo {author} {\bibfnamefont {G.}~\bibnamefont
  {Engelhardt}}, \bibinfo {author} {\bibfnamefont {M.}~\bibnamefont {Benito}},
  \bibinfo {author} {\bibfnamefont {G.}~\bibnamefont {Platero}}, \ and\
  \bibinfo {author} {\bibfnamefont {T.}~\bibnamefont {Brandes}},\ }\bibfield
  {title} {\enquote {\bibinfo {title} {Topologically enforced bifurcations in
  superconducting circuits},}\ }\href {\doibase 10.1103/PhysRevLett.118.197702}
  {\bibfield  {journal} {\bibinfo  {journal} {Phys. Rev. Lett.}\ }\textbf
  {\bibinfo {volume} {118}},\ \bibinfo {pages} {197702} (\bibinfo {year}
  {2017})}\BibitemShut {NoStop}%
\bibitem [{\citenamefont {Bastidas}\ \emph {et~al.}(2012)\citenamefont
  {Bastidas}, \citenamefont {Emary}, \citenamefont {Regler},\ and\
  \citenamefont {Brandes}}]{Bastidas2012}%
  \BibitemOpen
  \bibfield  {author} {\bibinfo {author} {\bibfnamefont {V.~M.}\ \bibnamefont
  {Bastidas}}, \bibinfo {author} {\bibfnamefont {C.}~\bibnamefont {Emary}},
  \bibinfo {author} {\bibfnamefont {B.}~\bibnamefont {Regler}}, \ and\ \bibinfo
  {author} {\bibfnamefont {T.}~\bibnamefont {Brandes}},\ }\bibfield  {title}
  {\enquote {\bibinfo {title} {Nonequilibrium quantum phase transitions in the
  {D}icke model},}\ }\href {\doibase 10.1103/PhysRevLett.108.043003} {\bibfield
   {journal} {\bibinfo  {journal} {Phys. Rev. Lett.}\ }\textbf {\bibinfo
  {volume} {108}},\ \bibinfo {pages} {043003} (\bibinfo {year}
  {2012})}\BibitemShut {NoStop}%
\bibitem [{\citenamefont {Engelhardt}\ \emph {et~al.}(2016)\citenamefont
  {Engelhardt}, \citenamefont {Benito}, \citenamefont {Platero},\ and\
  \citenamefont {Brandes}}]{Engelhardt2016}%
  \BibitemOpen
  \bibfield  {author} {\bibinfo {author} {\bibfnamefont {G.}~\bibnamefont
  {Engelhardt}}, \bibinfo {author} {\bibfnamefont {M.}~\bibnamefont {Benito}},
  \bibinfo {author} {\bibfnamefont {G.}~\bibnamefont {Platero}}, \ and\
  \bibinfo {author} {\bibfnamefont {T.}~\bibnamefont {Brandes}},\ }\bibfield
  {title} {\enquote {\bibinfo {title} {Topological instabilities in ac-driven
  bosonic systems},}\ }\href {\doibase 10.1103/PhysRevLett.117.045302}
  {\bibfield  {journal} {\bibinfo  {journal} {Phys. Rev. Lett.}\ }\textbf
  {\bibinfo {volume} {117}},\ \bibinfo {pages} {045302} (\bibinfo {year}
  {2016})}\BibitemShut {NoStop}%
\bibitem [{\citenamefont {Engelhardt}\ \emph {et~al.}(2013)\citenamefont
  {Engelhardt}, \citenamefont {Bastidas}, \citenamefont {Emary},\ and\
  \citenamefont {Brandes}}]{Engelhardt2013}%
  \BibitemOpen
  \bibfield  {author} {\bibinfo {author} {\bibfnamefont {G.}~\bibnamefont
  {Engelhardt}}, \bibinfo {author} {\bibfnamefont {V.~M.}\ \bibnamefont
  {Bastidas}}, \bibinfo {author} {\bibfnamefont {C.}~\bibnamefont {Emary}}, \
  and\ \bibinfo {author} {\bibfnamefont {T.}~\bibnamefont {Brandes}},\
  }\bibfield  {title} {\enquote {\bibinfo {title} {{ac-driven quantum phase
  transition in the Lipkin-Meshkov-Glick model}},}\ }\href {\doibase
  10.1103/PhysRevE.87.052110} {\bibfield  {journal} {\bibinfo  {journal} {Phys.
  Rev. E}\ }\textbf {\bibinfo {volume} {87}},\ \bibinfo {pages} {052110}
  (\bibinfo {year} {2013})}\BibitemShut {NoStop}%
\bibitem [{\citenamefont {Liu}\ \emph {et~al.}(2017)\citenamefont {Liu},
  \citenamefont {Hsieh},\ and\ \citenamefont {Cao}}]{Liu2017a}%
  \BibitemOpen
  \bibfield  {author} {\bibinfo {author} {\bibfnamefont {J.}~\bibnamefont
  {Liu}}, \bibinfo {author} {\bibfnamefont {C.-Y.}\ \bibnamefont {Hsieh}}, \
  and\ \bibinfo {author} {\bibfnamefont {J.}~\bibnamefont {Cao}},\ }\bibfield
  {title} {\enquote {\bibinfo {title} {Efficiency at maximum power of a quantum
  carnot engine with temperature tunable baths},}\ }\href@noop {} {\bibfield
  {journal} {\bibinfo  {journal} {arXiv:1710.06565}\ } (\bibinfo {year}
  {2017})}\BibitemShut {NoStop}%
\bibitem [{\citenamefont {Platero}\ and\ \citenamefont
  {Aguado}(2004)}]{Platero2004}%
  \BibitemOpen
  \bibfield  {author} {\bibinfo {author} {\bibfnamefont {G.}~\bibnamefont
  {Platero}}\ and\ \bibinfo {author} {\bibfnamefont {R.}~\bibnamefont
  {Aguado}},\ }\bibfield  {title} {\enquote {\bibinfo {title} {Photon-assisted
  transport in semiconductor nanostructures},}\ }\href@noop {} {\bibfield
  {journal} {\bibinfo  {journal} {Physics Reports}\ }\textbf {\bibinfo {volume}
  {395}},\ \bibinfo {pages} {1--157} (\bibinfo {year} {2004})}\BibitemShut
  {NoStop}%
\bibitem [{\citenamefont {Bavli}\ and\ \citenamefont
  {Metiu}(1993)}]{Bavli1993}%
  \BibitemOpen
  \bibfield  {author} {\bibinfo {author} {\bibfnamefont {R.}~\bibnamefont
  {Bavli}}\ and\ \bibinfo {author} {\bibfnamefont {H.}~\bibnamefont {Metiu}},\
  }\bibfield  {title} {\enquote {\bibinfo {title} {Properties of an electron in
  a quantum double well driven by a strong laser: {L}ocalization,
  low-frequency, and even-harmonic generation},}\ }\href {\doibase
  10.1103/PhysRevA.47.3299} {\bibfield  {journal} {\bibinfo  {journal} {Phys.
  Rev. A}\ }\textbf {\bibinfo {volume} {47}},\ \bibinfo {pages} {3299--3310}
  (\bibinfo {year} {1993})}\BibitemShut {NoStop}%
\bibitem [{\citenamefont {Dakhnovskii}\ and\ \citenamefont
  {Bavli}(1993)}]{Dakhnovskii1993}%
  \BibitemOpen
  \bibfield  {author} {\bibinfo {author} {\bibfnamefont {Y.}~\bibnamefont
  {Dakhnovskii}}\ and\ \bibinfo {author} {\bibfnamefont {R.}~\bibnamefont
  {Bavli}},\ }\bibfield  {title} {\enquote {\bibinfo {title} {Emission spectrum
  and localization of electrons in quantum-well systems induced by a strong
  laser field},}\ }\href {\doibase 10.1103/PhysRevB.48.11020} {\bibfield
  {journal} {\bibinfo  {journal} {Phys. Rev. B}\ }\textbf {\bibinfo {volume}
  {48}},\ \bibinfo {pages} {11020--11023} (\bibinfo {year} {1993})}\BibitemShut
  {NoStop}%
\bibitem [{\citenamefont {Graves}\ and\ \citenamefont
  {Gardiner}(1989)}]{Graves1989}%
  \BibitemOpen
  \bibfield  {author} {\bibinfo {author} {\bibfnamefont {P.R.}\ \bibnamefont
  {Graves}}\ and\ \bibinfo {author} {\bibfnamefont {D.}~\bibnamefont
  {Gardiner}},\ }\href@noop {} {\emph {\bibinfo {title} {Practical Raman
  spectroscopy}}}\ (\bibinfo  {publisher} {Springer},\ \bibinfo {year}
  {1989})\BibitemShut {NoStop}%
\bibitem [{\citenamefont {Mann}\ \emph {et~al.}(2017)\citenamefont {Mann},
  \citenamefont {Bakhtiari}, \citenamefont {Massel}, \citenamefont {Pelster},\
  and\ \citenamefont {Thorwart}}]{Mann2017}%
  \BibitemOpen
  \bibfield  {author} {\bibinfo {author} {\bibfnamefont {N.}~\bibnamefont
  {Mann}}, \bibinfo {author} {\bibfnamefont {M.~Reza}\ \bibnamefont
  {Bakhtiari}}, \bibinfo {author} {\bibfnamefont {F.}~\bibnamefont {Massel}},
  \bibinfo {author} {\bibfnamefont {A.}~\bibnamefont {Pelster}}, \ and\
  \bibinfo {author} {\bibfnamefont {M.}~\bibnamefont {Thorwart}},\ }\bibfield
  {title} {\enquote {\bibinfo {title} {Driven bose-hubbard model with a
  parametrically modulated harmonic trap},}\ }\href {\doibase
  10.1103/PhysRevA.95.043604} {\bibfield  {journal} {\bibinfo  {journal} {Phys.
  Rev. A}\ }\textbf {\bibinfo {volume} {95}},\ \bibinfo {pages} {043604}
  (\bibinfo {year} {2017})}\BibitemShut {NoStop}%
\bibitem [{\citenamefont {Komnik}\ and\ \citenamefont
  {Thorwart}(2016)}]{Komnik2016}%
  \BibitemOpen
  \bibfield  {author} {\bibinfo {author} {\bibfnamefont {A.}~\bibnamefont
  {Komnik}}\ and\ \bibinfo {author} {\bibfnamefont {M.}~\bibnamefont
  {Thorwart}},\ }\bibfield  {title} {\enquote {\bibinfo {title} {{BCS theory of
  driven superconductivity}},}\ }\href@noop {} {\bibfield  {journal} {\bibinfo
  {journal} {Eur. Phys. J. B}\ }\textbf {\bibinfo {volume} {89}},\ \bibinfo
  {pages} {244} (\bibinfo {year} {2016})}\BibitemShut {NoStop}%
\bibitem [{\citenamefont {Grossmann}\ \emph {et~al.}(1991)\citenamefont
  {Grossmann}, \citenamefont {Dittrich}, \citenamefont {Jung},\ and\
  \citenamefont {H\"anggi}}]{Grossmann1991}%
  \BibitemOpen
  \bibfield  {author} {\bibinfo {author} {\bibfnamefont {F.}~\bibnamefont
  {Grossmann}}, \bibinfo {author} {\bibfnamefont {T.}~\bibnamefont {Dittrich}},
  \bibinfo {author} {\bibfnamefont {P.}~\bibnamefont {Jung}}, \ and\ \bibinfo
  {author} {\bibfnamefont {P.}~\bibnamefont {H\"anggi}},\ }\bibfield  {title}
  {\enquote {\bibinfo {title} {Coherent destruction of tunneling},}\ }\href
  {\doibase 10.1103/PhysRevLett.67.516} {\bibfield  {journal} {\bibinfo
  {journal} {Phys. Rev. Lett.}\ }\textbf {\bibinfo {volume} {67}},\ \bibinfo
  {pages} {516--519} (\bibinfo {year} {1991})}\BibitemShut {NoStop}%
\bibitem [{\citenamefont {Mollow}(1969)}]{Mollow1969}%
  \BibitemOpen
  \bibfield  {author} {\bibinfo {author} {\bibfnamefont {B.~R.}\ \bibnamefont
  {Mollow}},\ }\bibfield  {title} {\enquote {\bibinfo {title} {Power spectrum
  of light scattered by two-level systems},}\ }\href {\doibase
  10.1103/PhysRev.188.1969} {\bibfield  {journal} {\bibinfo  {journal} {Phys.
  Rev.}\ }\textbf {\bibinfo {volume} {188}},\ \bibinfo {pages} {1969--1975}
  (\bibinfo {year} {1969})}\BibitemShut {NoStop}%
\bibitem [{\citenamefont {Shirai}\ \emph {et~al.}(2015)\citenamefont {Shirai},
  \citenamefont {Mori},\ and\ \citenamefont {Miyashita}}]{Shirai2015}%
  \BibitemOpen
  \bibfield  {author} {\bibinfo {author} {\bibfnamefont {T.}~\bibnamefont
  {Shirai}}, \bibinfo {author} {\bibfnamefont {T.}~\bibnamefont {Mori}}, \ and\
  \bibinfo {author} {\bibfnamefont {S.}~\bibnamefont {Miyashita}},\ }\bibfield
  {title} {\enquote {\bibinfo {title} {{Condition for emergence of the
  Floquet-Gibbs state in periodically driven open systems}},}\ }\href {\doibase
  10.1103/PhysRevE.91.030101} {\bibfield  {journal} {\bibinfo  {journal} {Phys.
  Rev. E}\ }\textbf {\bibinfo {volume} {91}},\ \bibinfo {pages} {030101}
  (\bibinfo {year} {2015})}\BibitemShut {NoStop}%
\bibitem [{\citenamefont {Shirai}\ \emph {et~al.}(2016)\citenamefont {Shirai},
  \citenamefont {Thingna}, \citenamefont {Mori}, \citenamefont {Denisov},
  \citenamefont {H\"anggi},\ and\ \citenamefont {Miyashita}}]{Shirai2016}%
  \BibitemOpen
  \bibfield  {author} {\bibinfo {author} {\bibfnamefont {T.}~\bibnamefont
  {Shirai}}, \bibinfo {author} {\bibfnamefont {J.}~\bibnamefont {Thingna}},
  \bibinfo {author} {\bibfnamefont {T.}~\bibnamefont {Mori}}, \bibinfo {author}
  {\bibfnamefont {S.}~\bibnamefont {Denisov}}, \bibinfo {author} {\bibfnamefont
  {P.}~\bibnamefont {H\"anggi}}, \ and\ \bibinfo {author} {\bibfnamefont
  {S.}~\bibnamefont {Miyashita}},\ }\bibfield  {title} {\enquote {\bibinfo
  {title} {{Effective Floquet–Gibbs states for dissipative quantum
  systems}},}\ }\href@noop {} {\bibfield  {journal} {\bibinfo  {journal} {New
  Journal of Physics}\ }\textbf {\bibinfo {volume} {18}},\ \bibinfo {pages}
  {053008} (\bibinfo {year} {2016})}\BibitemShut {NoStop}%
\bibitem [{\citenamefont {Frasca}(2003)}]{Frasca2003}%
  \BibitemOpen
  \bibfield  {author} {\bibinfo {author} {\bibfnamefont {M.}~\bibnamefont
  {Frasca}},\ }\bibfield  {title} {\enquote {\bibinfo {title} {Perturbative
  results on localization for a driven two-level system},}\ }\href {\doibase
  10.1103/PhysRevB.68.165315} {\bibfield  {journal} {\bibinfo  {journal} {Phys.
  Rev. B}\ }\textbf {\bibinfo {volume} {68}},\ \bibinfo {pages} {165315}
  (\bibinfo {year} {2003})}\BibitemShut {NoStop}%
\bibitem [{\citenamefont {Stockburger}(1999)}]{Stockburger1999}%
  \BibitemOpen
  \bibfield  {author} {\bibinfo {author} {\bibfnamefont {J.~T.}\ \bibnamefont
  {Stockburger}},\ }\bibfield  {title} {\enquote {\bibinfo {title} {Stabilizing
  coherent destruction of tunneling},}\ }\href {\doibase
  10.1103/PhysRevE.59.R4709} {\bibfield  {journal} {\bibinfo  {journal} {Phys.
  Rev. E}\ }\textbf {\bibinfo {volume} {59}},\ \bibinfo {pages} {R4709--R4712}
  (\bibinfo {year} {1999})}\BibitemShut {NoStop}%
\bibitem [{\citenamefont {Barata}\ and\ \citenamefont
  {Wreszinski}(2000)}]{Barata2000}%
  \BibitemOpen
  \bibfield  {author} {\bibinfo {author} {\bibfnamefont {J.~C.~A.}\
  \bibnamefont {Barata}}\ and\ \bibinfo {author} {\bibfnamefont {W.~F.}\
  \bibnamefont {Wreszinski}},\ }\bibfield  {title} {\enquote {\bibinfo {title}
  {Strong-coupling theory of two-level atoms in periodic fields},}\ }\href
  {\doibase 10.1103/PhysRevLett.84.2112} {\bibfield  {journal} {\bibinfo
  {journal} {Phys. Rev. Lett.}\ }\textbf {\bibinfo {volume} {84}},\ \bibinfo
  {pages} {2112--2115} (\bibinfo {year} {2000})}\BibitemShut {NoStop}%
\bibitem [{\citenamefont {Pigeau}\ \emph {et~al.}(2015)\citenamefont {Pigeau},
  \citenamefont {Rohr}, \citenamefont {De~Lepinay}, \citenamefont {Gloppe},
  \citenamefont {Jacques},\ and\ \citenamefont {Arcizet}}]{Pigeau2015}%
  \BibitemOpen
  \bibfield  {author} {\bibinfo {author} {\bibfnamefont {B.}~\bibnamefont
  {Pigeau}}, \bibinfo {author} {\bibfnamefont {S.}~\bibnamefont {Rohr}},
  \bibinfo {author} {\bibfnamefont {L.~M.}\ \bibnamefont {De~Lepinay}},
  \bibinfo {author} {\bibfnamefont {A.}~\bibnamefont {Gloppe}}, \bibinfo
  {author} {\bibfnamefont {V.}~\bibnamefont {Jacques}}, \ and\ \bibinfo
  {author} {\bibfnamefont {O.}~\bibnamefont {Arcizet}},\ }\bibfield  {title}
  {\enquote {\bibinfo {title} {{Observation of a phononic Mollow triplet in a
  multimode hybrid spin-nanomechanical system}},}\ }\href@noop {} {\bibfield
  {journal} {\bibinfo  {journal} {Nature Communications}\ }\textbf {\bibinfo
  {volume} {6}},\ \bibinfo {pages} {8603--8603} (\bibinfo {year}
  {2015})}\BibitemShut {NoStop}%
\bibitem [{\citenamefont {Yan}\ \emph {et~al.}(2016)\citenamefont {Yan},
  \citenamefont {Lu},\ and\ \citenamefont {Zheng}}]{Yan2016}%
  \BibitemOpen
  \bibfield  {author} {\bibinfo {author} {\bibfnamefont {Y.}~\bibnamefont
  {Yan}}, \bibinfo {author} {\bibfnamefont {Z.}~\bibnamefont {Lu}}, \ and\
  \bibinfo {author} {\bibfnamefont {H.}~\bibnamefont {Zheng}},\ }\bibfield
  {title} {\enquote {\bibinfo {title} {Resonance fluorescence of strongly
  driven two-level system coupled to multiple dissipative reservoirs},}\
  }\href@noop {} {\bibfield  {journal} {\bibinfo  {journal} {Annals of
  Physics}\ }\textbf {\bibinfo {volume} {371}},\ \bibinfo {pages} {159--182}
  (\bibinfo {year} {2016})}\BibitemShut {NoStop}%
\bibitem [{\citenamefont {Duan}\ \emph {et~al.}()\citenamefont {Duan},
  \citenamefont {Hsieh}, \citenamefont {J.},\ and\ \citenamefont {Cao}}]{Duan}%
  \BibitemOpen
  \bibfield  {author} {\bibinfo {author} {\bibfnamefont {C.}~\bibnamefont
  {Duan}}, \bibinfo {author} {\bibfnamefont {C.-Y.}\ \bibnamefont {Hsieh}},
  \bibinfo {author} {\bibfnamefont {Liu}\ \bibnamefont {J.}}, \ and\ \bibinfo
  {author} {\bibfnamefont {J.}~\bibnamefont {Cao}},\ }\href@noop {} {\enquote
  {\bibinfo {title} {Unusual transport properties with non-commutative
  system-bath coupling operators},}\ }\bibinfo {note} {In
  preparation}\BibitemShut {NoStop}%
\bibitem [{\citenamefont {Castro~Neto}\ \emph {et~al.}(2003)\citenamefont
  {Castro~Neto}, \citenamefont {Novais}, \citenamefont {Borda}, \citenamefont
  {Zar\'and},\ and\ \citenamefont {Affleck}}]{CastroNeto2003}%
  \BibitemOpen
  \bibfield  {author} {\bibinfo {author} {\bibfnamefont {A.~H.}\ \bibnamefont
  {Castro~Neto}}, \bibinfo {author} {\bibfnamefont {E.}~\bibnamefont {Novais}},
  \bibinfo {author} {\bibfnamefont {L.}~\bibnamefont {Borda}}, \bibinfo
  {author} {\bibfnamefont {Gergely}\ \bibnamefont {Zar\'and}}, \ and\ \bibinfo
  {author} {\bibfnamefont {I.}~\bibnamefont {Affleck}},\ }\bibfield  {title}
  {\enquote {\bibinfo {title} {Quantum magnetic impurities in magnetically
  ordered systems},}\ }\href {\doibase 10.1103/PhysRevLett.91.096401}
  {\bibfield  {journal} {\bibinfo  {journal} {Phys. Rev. Lett.}\ }\textbf
  {\bibinfo {volume} {91}},\ \bibinfo {pages} {096401} (\bibinfo {year}
  {2003})}\BibitemShut {NoStop}%
\bibitem [{\citenamefont {Guo}\ \emph {et~al.}(2012)\citenamefont {Guo},
  \citenamefont {Weichselbaum}, \citenamefont {von Delft},\ and\ \citenamefont
  {Vojta}}]{Guo2012}%
  \BibitemOpen
  \bibfield  {author} {\bibinfo {author} {\bibfnamefont {Cheng}\ \bibnamefont
  {Guo}}, \bibinfo {author} {\bibfnamefont {Andreas}\ \bibnamefont
  {Weichselbaum}}, \bibinfo {author} {\bibfnamefont {Jan}\ \bibnamefont {von
  Delft}}, \ and\ \bibinfo {author} {\bibfnamefont {Matthias}\ \bibnamefont
  {Vojta}},\ }\bibfield  {title} {\enquote {\bibinfo {title} {Critical and
  strong-coupling phases in one- and two-bath spin-boson models},}\ }\href
  {\doibase 10.1103/PhysRevLett.108.160401} {\bibfield  {journal} {\bibinfo
  {journal} {Phys. Rev. Lett.}\ }\textbf {\bibinfo {volume} {108}},\ \bibinfo
  {pages} {160401} (\bibinfo {year} {2012})}\BibitemShut {NoStop}%
\bibitem [{\citenamefont {Kohler}\ \emph {et~al.}(2013)\citenamefont {Kohler},
  \citenamefont {Hackl},\ and\ \citenamefont {Kehrein}}]{Kohler2013}%
  \BibitemOpen
  \bibfield  {author} {\bibinfo {author} {\bibfnamefont {Heiner}\ \bibnamefont
  {Kohler}}, \bibinfo {author} {\bibfnamefont {Andreas}\ \bibnamefont {Hackl}},
  \ and\ \bibinfo {author} {\bibfnamefont {Stefan}\ \bibnamefont {Kehrein}},\
  }\bibfield  {title} {\enquote {\bibinfo {title} {Nonequilibrium dynamics of a
  system with quantum frustration},}\ }\href {\doibase
  10.1103/PhysRevB.88.205122} {\bibfield  {journal} {\bibinfo  {journal} {Phys.
  Rev. B}\ }\textbf {\bibinfo {volume} {88}},\ \bibinfo {pages} {205122}
  (\bibinfo {year} {2013})}\BibitemShut {NoStop}%
\bibitem [{\citenamefont {Bruognolo}\ \emph {et~al.}(2014)\citenamefont
  {Bruognolo}, \citenamefont {Weichselbaum}, \citenamefont {Guo}, \citenamefont
  {von Delft}, \citenamefont {Schneider},\ and\ \citenamefont
  {Vojta}}]{Bruognolo2014}%
  \BibitemOpen
  \bibfield  {author} {\bibinfo {author} {\bibfnamefont {Benedikt}\
  \bibnamefont {Bruognolo}}, \bibinfo {author} {\bibfnamefont {Andreas}\
  \bibnamefont {Weichselbaum}}, \bibinfo {author} {\bibfnamefont {Cheng}\
  \bibnamefont {Guo}}, \bibinfo {author} {\bibfnamefont {Jan}\ \bibnamefont
  {von Delft}}, \bibinfo {author} {\bibfnamefont {Imke}\ \bibnamefont
  {Schneider}}, \ and\ \bibinfo {author} {\bibfnamefont {Matthias}\
  \bibnamefont {Vojta}},\ }\bibfield  {title} {\enquote {\bibinfo {title}
  {Two-bath spin-boson model: Phase diagram and critical properties},}\ }\href
  {\doibase 10.1103/PhysRevB.90.245130} {\bibfield  {journal} {\bibinfo
  {journal} {Phys. Rev. B}\ }\textbf {\bibinfo {volume} {90}},\ \bibinfo
  {pages} {245130} (\bibinfo {year} {2014})}\BibitemShut {NoStop}%
\bibitem [{\citenamefont {Shirley}(1965)}]{Shirley1965}%
  \BibitemOpen
  \bibfield  {author} {\bibinfo {author} {\bibfnamefont {J.~H.}\ \bibnamefont
  {Shirley}},\ }\bibfield  {title} {\enquote {\bibinfo {title} {Solution of the
  {S}chr\"odinger equation with a {H}amiltonian periodic in time},}\ }\href
  {\doibase 10.1103/PhysRev.138.B979} {\bibfield  {journal} {\bibinfo
  {journal} {Phys. Rev.}\ }\textbf {\bibinfo {volume} {138}},\ \bibinfo {pages}
  {B979--B987} (\bibinfo {year} {1965})}\BibitemShut {NoStop}%
\bibitem [{\citenamefont {Grifoni}\ and\ \citenamefont
  {H{\"a}nggi}(1998)}]{Grifoni1998}%
  \BibitemOpen
  \bibfield  {author} {\bibinfo {author} {\bibfnamefont {M.}~\bibnamefont
  {Grifoni}}\ and\ \bibinfo {author} {\bibfnamefont {P.}~\bibnamefont
  {H{\"a}nggi}},\ }\bibfield  {title} {\enquote {\bibinfo {title} {Driven
  quantum tunneling},}\ }\href {\doibase 10.1016/S0370-1573(98)00022-2}
  {\bibfield  {journal} {\bibinfo  {journal} {Physics Reports}\ }\textbf
  {\bibinfo {volume} {304}},\ \bibinfo {pages} {229--354} (\bibinfo {year}
  {1998})}\BibitemShut {NoStop}%
\bibitem [{\citenamefont {Carreno}\ \emph {et~al.}(2017)\citenamefont
  {Carreno}, \citenamefont {Valle},\ and\ \citenamefont
  {Laussy}}]{Carreno2017}%
  \BibitemOpen
  \bibfield  {author} {\bibinfo {author} {\bibfnamefont {J.~C.~L.}\
  \bibnamefont {Carreno}}, \bibinfo {author} {\bibfnamefont {E.~D.}\
  \bibnamefont {Valle}}, \ and\ \bibinfo {author} {\bibfnamefont {F.~P.}\
  \bibnamefont {Laussy}},\ }\bibfield  {title} {\enquote {\bibinfo {title}
  {Photon correlations from the {M}ollow triplet},}\ }\href@noop {} {\bibfield
  {journal} {\bibinfo  {journal} {Laser \& Photonics Reviews}\ }\textbf
  {\bibinfo {volume} {11}},\ \bibinfo {pages} {1700090} (\bibinfo {year}
  {2017})}\BibitemShut {NoStop}%
\bibitem [{par()}]{parityNote}%
  \BibitemOpen
  \href@noop {} {}\bibinfo {note} {Due to $\sigma_{\rm z} H_{\rm s}(t)
  \sigma_{\rm z} = H_{\rm s}(t+\tau/2) $, the Floquet states fulfill
  $\sigma_{\rm z} \left|\phi_{\lambda}(t) \right> = u_{\lambda}
  \left|\phi_{\lambda+\tau/2}(t) \right> $ with $u_{\lambda}\in \left\lbrace
  -1,1\right\rbrace$ and $u_0\neq u_1$. Using this in Eq.~(3) shows that
  $A_{\lambda\leftarrow\mu}^{(0)}=0$ for $\lambda\neq \mu$.}\BibitemShut
  {Stop}%
\bibitem [{\citenamefont {Gong}\ \emph {et~al.}(2009)\citenamefont {Gong},
  \citenamefont {Moralesmolina},\ and\ \citenamefont {H\"anggi}}]{Gong2009}%
  \BibitemOpen
  \bibfield  {author} {\bibinfo {author} {\bibfnamefont {J.}~\bibnamefont
  {Gong}}, \bibinfo {author} {\bibfnamefont {L.}~\bibnamefont {Moralesmolina}},
  \ and\ \bibinfo {author} {\bibfnamefont {P.}~\bibnamefont {H\"anggi}},\
  }\bibfield  {title} {\enquote {\bibinfo {title} {Many-body coherent
  destruction of tunneling},}\ }\href@noop {} {\bibfield  {journal} {\bibinfo
  {journal} {Phys. Rev. Lett.}\ }\textbf {\bibinfo {volume} {103}},\ \bibinfo
  {pages} {133002} (\bibinfo {year} {2009})}\BibitemShut {NoStop}%
\bibitem [{\citenamefont {Mananga}\ and\ \citenamefont
  {Charpentier}(2011)}]{Mananga2011}%
  \BibitemOpen
  \bibfield  {author} {\bibinfo {author} {\bibfnamefont {E.~S.}\ \bibnamefont
  {Mananga}}\ and\ \bibinfo {author} {\bibfnamefont {T.}~\bibnamefont
  {Charpentier}},\ }\bibfield  {title} {\enquote {\bibinfo {title}
  {{Introduction of the Floquet-Magnus expansion in solid-state nuclear
  magnetic resonance spectroscopy}},}\ }\href@noop {} {\bibfield  {journal}
  {\bibinfo  {journal} {Journal of Chemical Physics}\ }\textbf {\bibinfo
  {volume} {135}},\ \bibinfo {pages} {044109} (\bibinfo {year}
  {2011})}\BibitemShut {NoStop}%
\bibitem [{\citenamefont {S\'anchez}\ \emph {et~al.}(2008)\citenamefont
  {S\'anchez}, \citenamefont {Platero},\ and\ \citenamefont
  {Brandes}}]{Sanchez2008}%
  \BibitemOpen
  \bibfield  {author} {\bibinfo {author} {\bibfnamefont {R.}~\bibnamefont
  {S\'anchez}}, \bibinfo {author} {\bibfnamefont {G.}~\bibnamefont {Platero}},
  \ and\ \bibinfo {author} {\bibfnamefont {T.}~\bibnamefont {Brandes}},\
  }\bibfield  {title} {\enquote {\bibinfo {title} {Resonance fluorescence in
  driven quantum dots: {E}lectron and photon correlations},}\ }\href {\doibase
  10.1103/PhysRevB.78.125308} {\bibfield  {journal} {\bibinfo  {journal} {Phys.
  Rev. B}\ }\textbf {\bibinfo {volume} {78}},\ \bibinfo {pages} {125308}
  (\bibinfo {year} {2008})}\BibitemShut {NoStop}%
\bibitem [{\citenamefont {Brandes}\ \emph {et~al.}(2004)\citenamefont
  {Brandes}, \citenamefont {Aguado},\ and\ \citenamefont
  {Platero}}]{Brandes2004}%
  \BibitemOpen
  \bibfield  {author} {\bibinfo {author} {\bibfnamefont {T.}~\bibnamefont
  {Brandes}}, \bibinfo {author} {\bibfnamefont {R.}~\bibnamefont {Aguado}}, \
  and\ \bibinfo {author} {\bibfnamefont {G.}~\bibnamefont {Platero}},\
  }\bibfield  {title} {\enquote {\bibinfo {title} {Charge transport through
  open driven two-level systems with dissipation},}\ }\href {\doibase
  10.1103/PhysRevB.69.205326} {\bibfield  {journal} {\bibinfo  {journal} {Phys.
  Rev. B}\ }\textbf {\bibinfo {volume} {69}},\ \bibinfo {pages} {205326}
  (\bibinfo {year} {2004})}\BibitemShut {NoStop}%
\bibitem [{\citenamefont {S\'anchez}\ \emph {et~al.}(2007)\citenamefont
  {S\'anchez}, \citenamefont {Platero},\ and\ \citenamefont
  {Brandes}}]{Sanchez2007}%
  \BibitemOpen
  \bibfield  {author} {\bibinfo {author} {\bibfnamefont {R.}~\bibnamefont
  {S\'anchez}}, \bibinfo {author} {\bibfnamefont {G.}~\bibnamefont {Platero}},
  \ and\ \bibinfo {author} {\bibfnamefont {T.}~\bibnamefont {Brandes}},\
  }\bibfield  {title} {\enquote {\bibinfo {title} {Resonance fluorescence in
  transport through quantum dots: {N}oise properties},}\ }\href {\doibase
  10.1103/PhysRevLett.98.146805} {\bibfield  {journal} {\bibinfo  {journal}
  {Phys. Rev. Lett.}\ }\textbf {\bibinfo {volume} {98}},\ \bibinfo {pages}
  {146805} (\bibinfo {year} {2007})}\BibitemShut {NoStop}%
\bibitem [{\citenamefont {Restrepo}\ \emph {et~al.}(2016)\citenamefont
  {Restrepo}, \citenamefont {Cerrillo}, \citenamefont {Bastidas}, \citenamefont
  {Angelakis},\ and\ \citenamefont {Brandes}}]{Restrepo2016}%
  \BibitemOpen
  \bibfield  {author} {\bibinfo {author} {\bibfnamefont {S.}~\bibnamefont
  {Restrepo}}, \bibinfo {author} {\bibfnamefont {J.}~\bibnamefont {Cerrillo}},
  \bibinfo {author} {\bibfnamefont {V.~M.}\ \bibnamefont {Bastidas}}, \bibinfo
  {author} {\bibfnamefont {D.~G.}\ \bibnamefont {Angelakis}}, \ and\ \bibinfo
  {author} {\bibfnamefont {T.}~\bibnamefont {Brandes}},\ }\bibfield  {title}
  {\enquote {\bibinfo {title} {Driven open quantum systems and {F}loquet
  stroboscopic dynamics},}\ }\href {\doibase 10.1103/PhysRevLett.117.250401}
  {\bibfield  {journal} {\bibinfo  {journal} {Phys. Rev. Lett.}\ }\textbf
  {\bibinfo {volume} {117}},\ \bibinfo {pages} {250401} (\bibinfo {year}
  {2016})}\BibitemShut {NoStop}%
\bibitem [{\citenamefont {Magazz\`u}\ \emph {et~al.}(2018)\citenamefont
  {Magazz\`u}, \citenamefont {Denisov},\ and\ \citenamefont
  {H\"anggi}}]{Magazzu2018}%
  \BibitemOpen
  \bibfield  {author} {\bibinfo {author} {\bibfnamefont {L.}~\bibnamefont
  {Magazz\`u}}, \bibinfo {author} {\bibfnamefont {S.}~\bibnamefont {Denisov}},
  \ and\ \bibinfo {author} {\bibfnamefont {P.}~\bibnamefont {H\"anggi}},\
  }\bibfield  {title} {\enquote {\bibinfo {title} {Asymptotic {F}loquet states
  of a periodically driven spin-boson system in the nonperturbative coupling
  regime},}\ }\href {\doibase 10.1103/PhysRevE.98.022111} {\bibfield  {journal}
  {\bibinfo  {journal} {Phys. Rev. E}\ }\textbf {\bibinfo {volume} {98}},\
  \bibinfo {pages} {022111} (\bibinfo {year} {2018})}\BibitemShut {NoStop}%
\bibitem [{\citenamefont {Xu}\ and\ \citenamefont {Cao}(2016)}]{Xu2016}%
  \BibitemOpen
  \bibfield  {author} {\bibinfo {author} {\bibfnamefont {D.}~\bibnamefont
  {Xu}}\ and\ \bibinfo {author} {\bibfnamefont {J.}~\bibnamefont {Cao}},\
  }\bibfield  {title} {\enquote {\bibinfo {title} {Non-canonical distribution
  and non-equilibrium transport beyond weak system-bath coupling regime: A
  polaron transformation approach},}\ }\href@noop {} {\bibfield  {journal}
  {\bibinfo  {journal} {Frontiers of Physics}\ }\textbf {\bibinfo {volume}
  {11}},\ \bibinfo {pages} {110308} (\bibinfo {year} {2016})}\BibitemShut
  {NoStop}%
\bibitem [{\citenamefont {Duan}\ \emph {et~al.}(2017)\citenamefont {Duan},
  \citenamefont {Tang}, \citenamefont {Cao},\ and\ \citenamefont
  {Wu}}]{Duan2017}%
  \BibitemOpen
  \bibfield  {author} {\bibinfo {author} {\bibfnamefont {C.}~\bibnamefont
  {Duan}}, \bibinfo {author} {\bibfnamefont {Z.}~\bibnamefont {Tang}}, \bibinfo
  {author} {\bibfnamefont {J.}~\bibnamefont {Cao}}, \ and\ \bibinfo {author}
  {\bibfnamefont {J.}~\bibnamefont {Wu}},\ }\bibfield  {title} {\enquote
  {\bibinfo {title} {Zero-temperature localization in a sub-ohmic spin-boson
  model investigated by an extended hierarchy equation of motion},}\ }\href
  {\doibase 10.1103/PhysRevB.95.214308} {\bibfield  {journal} {\bibinfo
  {journal} {Phys. Rev. B}\ }\textbf {\bibinfo {volume} {95}},\ \bibinfo
  {pages} {214308} (\bibinfo {year} {2017})}\BibitemShut {NoStop}%
\bibitem [{\citenamefont {Lee}\ \emph {et~al.}(2012)\citenamefont {Lee},
  \citenamefont {Cao},\ and\ \citenamefont {Gong}}]{Lee2012}%
  \BibitemOpen
  \bibfield  {author} {\bibinfo {author} {\bibfnamefont {C.~K.}\ \bibnamefont
  {Lee}}, \bibinfo {author} {\bibfnamefont {J.}~\bibnamefont {Cao}}, \ and\
  \bibinfo {author} {\bibfnamefont {J.}~\bibnamefont {Gong}},\ }\bibfield
  {title} {\enquote {\bibinfo {title} {Noncanonical statistics of a spin-boson
  model: Theory and exact monte carlo simulations},}\ }\href {\doibase
  10.1103/PhysRevE.86.021109} {\bibfield  {journal} {\bibinfo  {journal} {Phys.
  Rev. E}\ }\textbf {\bibinfo {volume} {86}},\ \bibinfo {pages} {021109}
  (\bibinfo {year} {2012})}\BibitemShut {NoStop}%
\end{thebibliography}%

\begin{widetext}
\begin{center}
\textbf
{\huge Supplementary information}
\end{center}

In this supplementary information we explicitly show how do derive the analytical expression for the Floquet states transition elements Eq.~\eqref{eq:rates} in the high-frequency regime, which is given in Eq.~\eqref{eq:analyticCoefficients}. According to  Floquet theory, the time-evolution operator can be  written as
\begin{equation}
	\hat U (t,0)=  e^{- i \hat \Lambda (t) } e^{- i H_{\rm eff} (t) } 
	\label{eq:timeEvolutionOperator}
\end{equation}
with a time-periodic hermitian operator $\hat \Lambda(t)$ and a Floquet Hamiltonian $ H_{\rm eff}  $ whose eigenvalues and eigenstates represent quasienergies and stroposcopic Floquet states respectively.

\section{Transformation into a rotating frame}

The  system Hamiltonian describing the isolated spin system  reads 
\begin{equation}
	H_{\rm s} = \frac{h_{\rm x}}{2}\sigma_{x} + \frac{h_{\rm z,1}}{2}\cos (\Omega t) \sigma_{z} ,
 \end{equation}
where we consider the case $h_{\rm z,0}=0$ for simplicity.
The  transformation into a rotating frame is defined by the unitary operator
\begin{align}
\hat U_{\rm rot} (t) &= \exp\left[- i  \sigma_{\rm z}   \theta(t)  \right] ,  \nonumber \\
\theta (t) &= \frac{h_{\rm z,1}}{2 \Omega} \sin(\Omega t  )   ,
\end{align}
which transforms the system wave function $\left|\Psi (t)\right> = \hat U_{\rm rot} ^\dagger (t)\left|\tilde \Psi (t)\right>$, with $\left|\tilde \Psi (t)\right>$ being the wave function in the rotating frame.

Evaluating the expressions of the unitary operators, we find
\begin{align}
	U_{\rm rot} (t) &= 1 \cdot  \cos \theta (t)  - i \sigma_{\rm z} \sin \theta (t) ,
\end{align}
which gives rise to the transformation of the Pauli operators
\begin{align}
	U_{\rm rot}^\dagger  (t)\sigma_{\rm z} U_{\rm rot}(t) &= \sigma_{\rm z} , \nonumber  \\
	U_{\rm rot}^\dagger  (t)\sigma_{\rm x} U_{\rm rot} (t) &= \left[ \cos^2(\theta ) - \sin^2(\theta )\right] \sigma_{\rm x}  + \cos(\theta ) \sin(\theta )i \left[\sigma_{\rm z} \sigma_{\rm x}   - \sigma_{\rm x} \sigma_{\rm z}  \right], \nonumber \\
	&= \cos(2 \theta ) \sigma_{\rm x} -  \sin(2 \theta ) \sigma_{\rm y}\nonumber \\
		U_{\rm rot}^\dagger  (t)\sigma_{\rm y} U_{rot}(t) &=  \cos(2 \theta ) \sigma_{\rm y} +  \sin(2 \theta) \sigma_{\rm x}.
\end{align}

\section{Rotating wave approximation}

This can be expanded as a Fourier decomposition
\begin{equation}
H_{\rm s} (t)= \sum_{n=-\infty}^{\infty} H_{\rm s}^{(n)} e^{in \Omega t},
\end{equation}%
at which each Fourier component reads
\begin{equation}
H_{\rm s}^{(n)} = \frac{h_x}{2} \mathcal J_{n} \left(\frac{ h_{\rm z,1}} {\Omega }\right)
\begin{cases} 
 \sigma_{\rm x }& n\; \text{even}\\
 i \sigma_{\rm y } & n \; \text{odd}
\end{cases},
\end{equation}
where we have used  the integral representation of the Bessel functions
\begin{equation}
J_{n}(x) = \frac{1}{2 \pi } \int_{-\pi}^{\pi} e^{x \sin \tau + n\tau} d\tau
\end{equation}
Following Ref.~\cite{Mananga2011} , a high-frequency expansion of the system time-evolution operator in Eq.~\eqref{eq:timeEvolutionOperator} reads
\begin{align}
   H_{\rm eff} &= H_{\rm s}^{(0)}  + \frac 12 \sum_{m\neq 0} \frac{1}{m \omega} \left[H_{\rm s}^{(m)},H_{\rm s}^{(-m) } \right] + \sum_{m\neq 0} \frac{1}{m \omega} \left[H_{\rm s}^{(0)},H_{\rm s}^{(-m) } \right]  +  \mathcal O \left(\frac{1}{\Omega} \right) ,\nonumber \\
   \hat \Lambda(t) &= \sum _{ n \neq 0 } \frac{1}{i n \Omega} H_{\rm s}^{(n)} (e^{i n \Omega t}-1 ) + \mathcal O \left(\frac{1}{\Omega^2} \right)
\end{align}
where we consider the lowest non-vanishing terms in order of $\frac{1}{\Omega}$. For the system under consideration, we find the well-known result
\begin{equation}
H_{\rm eff} =  \frac { h_{\rm x}}  2 \mathcal J_{ 0}( h_{\rm z,1}/ \Omega ) \sigma_{x} - 2 \frac{h_{x}^2}{\Omega}\sum_{k>0} \frac{ \mathcal J_{2k -1 } ( h_{\rm z,1}/ \Omega ) } {2k -1 } \sigma_{\rm z}  + \mathcal O \left(\frac{1}{\Omega} \right)
\end{equation} 
 and
 \begin{align}
 \hat \Lambda(t) &=  \frac{ h_{\rm x}}{2} \sum _{k=1}^{\infty} \frac{1}{  2 k \Omega} \mathcal J_{ 2k} \left(\frac{ h_{\rm z,1}}{\Omega} \right)  2 \sin(2k \Omega t) \sigma_{\rm x} + \frac{ h_{\rm x}}{2}\sum _{k=1}^{\infty} \frac{1}{  (2 k-1) \Omega} \mathcal J_{ (2 k-1) } \left(\frac{ h_{\rm z,1}}{\Omega} \right)  2 (\cos((2k-1) \Omega t) -1) \sigma_{\rm y} +  \mathcal O\left(\frac{1}{\Omega^2} \right)\nonumber \\
 &=  \frac{ h_{\rm x} }{ \Omega}   \sum _{k=1}^{\infty}  l_{2k}   \sin(2 k \Omega t) \sigma_{\rm x} +   \frac{h_{\rm x} }{ \Omega}   \sum _{k=1}^{\infty}  l_{2k-1}   (\cos((2k-1) \Omega t) -1) \sigma_{\rm y} +  \mathcal O \left(\frac{1}{\Omega^2} \right)\nonumber \\
 &=  \frac{h_{\rm x} }{\Omega} l_{\rm x}(t) \sigma_{\rm x}  +  \frac{ h_{\rm x} }{\Omega}  l_{\rm y}(t)\sigma_{\rm y}  +  \mathcal O \left(\frac{1}{\Omega^2} \right),
 \end{align}
 where $l_n (z)= \frac{1}{  n\Omega} \mathcal J_{ n} \left(z \right)$.
We use this to calculate the terms $a_{\mu,\lambda}^{(n)}$
\begin{align}
a_{\mu,\lambda}^{(n)} &= \frac1\tau \int_{0}^{\tau} \left< u_{\mu}(t)\right| \hat \sigma_{\theta}  \left| u_{\lambda}(t)\right>e^{- i n \Omega t} dt\nonumber \\
&= \frac1\tau \int_{0}^{\tau} \left< u_{\mu}(0)\right| e^{ i \hat \Lambda (t)  }  U_{\rm rot}^\dagger  \left[g_x \sigma_x+ g_z \sigma_z \right] U_{\rm rot} e^{ -i\hat \Lambda( t)   }  \left| u_{\lambda}(0)\right>  e^{- i n \Omega t} dt 
\end{align}
Using the expansion
\begin{equation}
e^{ -i \hat \Lambda (t)   } \approx 1  -i\hat \Lambda (t),
\end{equation}
we can explicitly solve the integral and find
\begin{equation}
a_{\mu,\lambda}^{(n)} = g_x a_{\rm x,\mu,\lambda}^{(n)} + g_{z} a_{\rm z,\mu,\lambda}^{(n)},
\end{equation}
with
\begin{align}
a_{x,\mu,\lambda}^{(n)}  &=   \frac1\tau \int_{0}^{\tau} \left< u_{\mu}(0)\right| e^{  i \hat \Lambda(t)   }   \left[ \cos(2 \theta ) \sigma_{\rm x} -  \sin(2 \theta ) \sigma_{\rm y} \right] e^{ - i \hat  \Lambda(t)   }  \left| u_{\lambda}(0)\right>  e^{- i m \Omega t} dt \nonumber \\ 
 &\approx \frac{ h_{\rm x}}{\Omega}\frac1\tau \int_{0}^{\tau} \left< u_{\mu}(0)\right|   
   \left[ \cos(2 \theta ) 
\left[  \sigma_{\rm x}  +  i \frac{ h_{\rm x}}{\Omega} l_{\rm y} (t) \left[\sigma_{\rm y} ,\sigma_{\rm x}  \right]  \right]
 -  \sin(2 \theta ) \left[  \sigma_{\rm y}  +  i \frac{ h_{\rm x}}{\Omega} l_{\rm x} (t) \left[\sigma_{\rm x} ,\sigma_{\rm y}  \right]  \right] \right] 
 \left| u_{\lambda}(0)\right>  e^{- i n \Omega t} dt \nonumber \\ 
 &= \frac{ h_{\rm x}}{\Omega} \frac1\tau \int_{0}^{\tau} \left< u_{\mu}(0)\right|   
    \cos(2 \theta ) 
 \sigma_{\rm x}  -  \sin(2 \theta ) 
 \sigma_{\rm y}  + 2 \frac{ h_{\rm x}}{\Omega} \left[ l_{\rm y}(t) \cos(2 \theta ) +  l_{\rm x} (t) \sin(2 \theta ) \right] \sigma_{\rm z}
 \left| u_{\lambda}(0)\right>  e^{- i n \Omega t} dt ,\nonumber \\
a_{\rm z,\mu,\lambda}^{(n)}  &=  \frac{ h_{\rm x}}{\Omega} \frac1\tau \int_{0}^{\tau} \left< u_{\mu}(0)\right| e^{ i \hat \Lambda(t)   }   \sigma_{\rm z} e^{  i \hat \Lambda(t)   }  \left| u_{\lambda}(0)\right>  e^{- i n \Omega t} dt \nonumber \\ 
 &\approx  \frac{ h_{\rm x}}{\Omega} \frac1\tau \int_{0}^{\tau} \left< u_{\mu}(0)\right|   \sigma_z +  i \frac{ h_{\rm x}}{\Omega} l_{\rm y} (t) \left[\sigma_{\rm y} ,\sigma_{\rm z}  \right] + i \frac{ h_{\rm x}}{\Omega} l_{\rm x} (t) \left[\sigma_{\rm x} ,\sigma_{\rm z}  \right]  
 \left| u_{\lambda}(0)\right>    e^{- i n \Omega t} dt  \nonumber \\
  &=  \frac{ h_{\rm x}}{\Omega}  \frac1\tau \int_{0}^{\tau} \left< u_{\mu}(0)\right|   
    \sigma_{z} - 2 \frac{ h_{\rm x}}{\Omega} l_{\rm y}(t)\sigma_x + 2 \frac{ h_{\rm x}}{\Omega} l_{\rm x}(t)\sigma_{\rm y}
 \left| u_{\lambda}(0)\right>  e^{- i n \Omega t} dt 
\end{align}
This integrals can be solved analytically, giving
\begin{align}
a_{x,\mu,\lambda}^{(n)}
 =&  \frac 12 
 \left[ \mathcal J_{-n} \left(  \frac{ h_{\rm z,1} }{\Omega} \right)+                      \mathcal J_{-n} \left(  -\frac{ h_{\rm z,1} }{\Omega} \right) 
 \right]
 \left< u_{\mu}(0)\right| \sigma_{\rm x}   \left| u_{\lambda}(0)\right>\nonumber \\
 &-  \frac 1{2i} 
 \left[ \mathcal J_{-n} \left(  \frac{ h_{\rm z,1} }{\Omega} \right) - \mathcal J_{-n} \left( - \frac{ h_{\rm z,1} }{\Omega} \right) 
 \right]
 \left< u_{\mu}(0)\right| \sigma_{\rm y}   \left| u_{\lambda}(0)\right>\nonumber \\
 &+\frac{ h_{x}}{\Omega} \mathcal A_{\rm x}^{(n)}  \left< u_{\mu}(0)\right| \sigma_{\rm z}   \left| u_{\lambda}(0)\right> \nonumber \\
 =& 
 \mathcal J_{n} \left(  \frac{ h_{\rm z,1} }{\Omega} \right)\delta_{n\,\text{mod}\,2,0}
 \left< \sigma_{x}  \right>_{\mu\lambda}
 -  i \mathcal J_{n} \left(  \frac{ h_{\rm z,1} }{\Omega} \right)\delta_{n\,\text{mod}\,2,1}
 \left< \sigma_{\rm y}  \right>_{\mu\lambda}
 + \mathcal A_{\rm x}^{(n)}  \left< \sigma_{\rm z}  \right>_{\mu\lambda},
\end{align}
and
\begin{align}
a_{\rm z,\mu,\lambda}^{(n)}
 =& \mathcal  B_{\rm x}^{(n)}
\left< \sigma_{\rm x}  \right>_{\mu\lambda} 
+\mathcal B_{\rm y}^{(n)}
 \left< \sigma_{\rm y}  \right>_{\mu\lambda}
 +  \delta_{n,0}  \left< \sigma_{\rm z}  \right>_{\mu\lambda},
\end{align}
where we have defined
\begin{align}
\mathcal A_{\rm x}^{(n)} &= \frac1\tau \int_{0}^{\tau}  2 \left[ l_{\rm y}(t)  \cos(2 \theta ) +  l_{\rm x}(t)  \sin(2 \theta ) \right] e^{- in \Omega t} dt 
=\mathcal A_{\rm x,1}^{(n)} + \mathcal A_{\rm x,2}^{(n)} 
\end{align}
Evaluating these terms we obtain
\begin{align}
\mathcal A_{\rm x,1}^{(n)}  &= \frac1\tau \int_{0}^{\tau}  2 \frac{ h_{\rm x}}{\Omega}  l_{\rm y}(t)  \cos(2 \theta )  e^{- i n \Omega t} dt
= 2 \sum _{k=1}^{\infty}\frac{ h_{\rm x}}{\Omega}  l_{2k-1}    \frac1\tau \int_{0}^{\tau}        \cos(2 \theta )  (\cos( ( (2k-1)   \Omega t) -1) e^{- i n \Omega t} dt\nonumber \\
&=  \sum _{k=1}^{\infty}\frac{ h_{\rm x}}{\Omega}  l_{2k-1} 
 \left[
 -\mathcal J_{-n} \left(   \frac{ h_{\rm z,1}}{\Omega} \right) 
  -  \mathcal     J_{-n} \left( -\frac{ h_{\rm z,1} }{\Omega} \right) \right.\nonumber \\ &\left.
    +\frac 12 \mathcal J_{-n+2k-1} \left(   \frac{ h_{\rm z,1}}{\Omega} \right) 
  +\frac 12  \mathcal     J_{-n+2k-1} \left(  - \frac{ h_{\rm z,1} }{\Omega} \right) \right.\nonumber \\ &\left.
  +\frac 12 \mathcal J_{-n-2k+1} \left(   \frac{ h_{\rm z,1} }{\Omega} \right) 
  + \frac 12  \mathcal     J_{-n-2k+1} \left( - \frac{ h_{\rm z,1} }{\Omega} \right) 
 \right]\nonumber \\
&=  \sum _{k=1}^{\infty}\frac{ h_{\rm x}}{\Omega}  l_{2k-1} 
 \left[
 -\mathcal J_{-n}  \left(  \frac{ h_{\rm z,1} }{\Omega} \right) \left( 1+ (-1)^{n} \right) 
  \right.\nonumber \\ &\left.
    +\frac 12 \mathcal J_{-n+2k-1} \left(   \frac{ h_{\rm z,1}}{\Omega} \right) \left( 1+ (-1)^{-n+2k -1} \right) 
   \right.\nonumber \\ &\left.
  +\frac 12 \mathcal J_{-n-2k+1} \left(   \frac{ h_{\rm z,1}}{\Omega} \right) 
  \left( 1+ (-1)^{-n-2k +1} \right)  
 \right]\nonumber \\
&=  \sum _{k=1}^{\infty}\frac{ h_{\rm x}}{\Omega}  l_{2k-1} 
 \left[
 -\mathcal J_{-n}  \left(  \frac{ h_{\rm z,1}}{\Omega} \right) \left( 1+ (-1)^{n} \right) 
  \right.\nonumber \\ &\left.
    +\frac 12 \mathcal J_{-n+2k-1} \left(   \frac{ h_{\rm z,1}}{\Omega} \right) \left( 1- (-1)^{-n} \right) 
   \right.\nonumber \\ &\left.
  -(-1)^{(2k-1)} \frac 12 \mathcal J_{-n-2k+1} \left(   \frac{ h_{\rm z,1} }{\Omega} \right) 
  \left( 1- (-1)^{-n} \right)  
 \right]
\end{align}
\begin{align}
\mathcal A_{\rm x,2}^{(n)}  &= \frac1\tau \int_{0}^{\tau}  2 \frac{ h_{\rm x}}{\Omega}  l_{\rm y}(t)  \sin(2 \theta )  e^{- i n \Omega t} dt
= 2 \sum _{k=1}^{\infty} l_{2k-1}  \frac1\tau \int_{0}^{\tau}        \sin(2 \theta )  \sin( 2k    \Omega t)  e^{- i n \Omega t} dt\nonumber \\
&=  - \sum _{k=1}^{\infty} l_{2k} 
 \left[
    +\frac 12 \mathcal J_{-n+2k} \left(   \frac{ h_{\rm z,1}}{\Omega} \right) 
  -\frac 12  \mathcal     J_{-n+2k} \left(  - \frac{ h_{\rm z,1} }{\Omega} \right) \right.\nonumber \\ &\left.
  -\frac 12 \mathcal J_{-n-2k} \left(   \frac{ h_{\rm z,1} }{\Omega} \right) 
  + \frac 12  \mathcal     J_{-n-2k} \left( - \frac{ h_{\rm z,1} }{\Omega} \right) 
 \right]\nonumber \\
&=  - \sum _{k=1}^{\infty} l_{2k} 
 \left[
    +\frac 12 \mathcal J_{-n+2k} \left(   \frac{ h_{\rm z,1}}{\Omega} \right) 
   \left( 1- (-1)^{-n+2k } \right)  \right.\nonumber \\ &\left.
  -\frac 12 \mathcal J_{-n-2k} \left(   \frac{ h_{\rm z,1} }{\Omega} \right) 
  \left( 1- (-1)^{-n-2k } \right) 
 \right]\nonumber \\
&=  - \sum _{k=1}^{\infty} l_{2k} 
 \left[
    +\frac 12 \mathcal J_{-n+2k} \left(   \frac{ h_{\rm z,1}}{\Omega} \right) 
   \left( 1- (-1)^{-n } \right) 
  -(-1)^{2k}\frac 12 \mathcal J_{-n-2k} \left(   \frac{ h_{\rm z,1} }{\Omega} \right) 
  \left( 1- (-1)^{-n } \right) 
 \right]
\end{align}

\begin{align}
\mathcal A_{1}^{(n)}  
&=  
 -\mathcal J_{n}  \left(  \frac{ h_{\rm z,1} }{\Omega} \right) \delta_{n \,\text{mod}\, 2, 0}\sum _{k=1}^{\infty} l_{2k-1}  \nonumber  \\
 &+\frac 12   \delta_{n \,\text{mod}\, 2, 1} \sum _{m=1}^{\infty} l_{m} 
 \left[
    \mathcal J_{-n+m} \left(   \frac{ h_{\rm z,1}}{\Omega} \right) 
  -(-1)^{m}  \mathcal J_{-n-m} \left(   \frac{ h_{\rm z,1} }{\Omega} \right)   
 \right]
\end{align}

\begin{align}
\mathcal B_{\rm x,1}^{(n)}  &= \frac1\tau \int_{0}^{\tau}  2 l_{\rm y}(t)   e^{- i n \Omega t} dt
= 2 \sum _{k=1}^{\infty} l_{2k-1}   \frac1\tau \int_{0}^{\tau}        (\cos( ( (2k-1)   \Omega t) -1) e^{- i n \Omega t}  dt  \nonumber \\
&=  \sum _{k=1}^{\infty} l_{2k-1} 
 \left[
 -2 \delta_{n,0} +  \delta_{2k-1-n,0} + \delta_{-2k-1-n,0}\right] 
 =  -2 \delta_{n,0} \sum _{k=1}^{\infty} l_{2k-1} + \delta_{n \text{mod} 2, 1} l_{\left|n\right|} 
\end{align}

\begin{align}
\mathcal B_{y,1}^{(n )}  &= \frac1\tau \int_{0}^{\tau} 2 l_{x}(t)   e^{- i n \Omega t} dt
= 2 \sum _{k=1}^{\infty} l_{2k}    \frac1\tau \int_{0}^{\tau}      \sin(k\Omega t) e^{- i n \Omega t}  dt  \nonumber \\
&=  \frac 1 i \sum _{k=1}^{\infty} l_{2k} 
 \left[
 \delta_{2k-n,0} - \delta_{-2k-n,0}\right] 
 = -i \delta_{n \text{mod} 2, 0} l_{\left|n\right|} 
\end{align}

\end{widetext}

\end{document}